\documentclass{ieeeaccess}
\usepackage{cite}
\usepackage{amsmath,amssymb,amsfonts}
\usepackage{algorithmic}
\usepackage{enumitem}
\usepackage{float}
\usepackage{makecell}
\usepackage{xcolor, soul}
\sethlcolor{yellow}

\usepackage{booktabs}
\usepackage{graphicx}
\usepackage{textcomp}
\def\BibTeX{{\rm B\kern-.05em{\sc i\kern-.025em b}\kern-.08em
    T\kern-.1667em\lower.7ex\hbox{E}\kern-.125emX}}
\begin{document}
\history{Date of publication xxxx 00, 0000, date of current version xxxx 00, 0000.}
\doi{10.1109/ACCESS.2022.11635}

\title{Underwater Acoustic Communication Channel Modeling using Reservoir Computing}
\author{\uppercase{Oluwaseyi Onasami}\authorrefmark{1},
\uppercase{Ming Feng}\authorrefmark{2}, 
\uppercase{Hao Xu}\authorrefmark{2}, 
\uppercase{Mulugeta Haile\authorrefmark{3}, and Lijun Qian}.\authorrefmark{1}}
\address[1]{Department of Electrical and Computer Engineering, Prairie View A\&M University, Prairie View, TX 77446, USA. (e-mail: oonasami@pvamu.edu, liqian@pvamu.edu)}
\address[2]{Department of Electrical and Biomedical Engineering, University of Nevada, Reno, NV 89557, USA. (e-mail: mingf@unr.edu, haoxu@unr.edu)}
\address[3]{U.S. Army Research Laboratory, Aberdeen Proving Ground, MD 21005, USA. (e-mail: mulugeta.a.haile.civ@army.mil)}
\tfootnote{This research work is supported in part by the U.S. Army Research Office (ARO) under agreement number W911NF-20-2-0266 and the U.S. Office of the Under Secretary of Defense for Research and Engineering (OUSD(R\&E)) under agreement number FA8750-15-2-0119. The U.S. Government is authorized to reproduce and distribute reprints for governmental purposes notwithstanding any copyright notation thereon. }

\markboth
{Oluwaseyi \headeretal: Underwater Acoustic Communication Channel Modeling using Reservoir Computing}
{Oluwaseyi \headeretal: Underwater Acoustic Communication Channel Modeling using Reservoir Computing}

\corresp{Corresponding author: Lijun Qian (e-mail: liqian@pvamu.edu).}

\begin{abstract}
Underwater acoustic (UWA) communications have been widely used but greatly impaired due to the complicated nature of the underwater environment. In order to improve UWA communications, modeling and understanding the UWA channel is indispensable. However, there exist many challenges due to the high uncertainties of the underwater environment and the lack of real-world measurement data. In this work, the capability of reservoir computing and deep learning has been explored for modeling the UWA communication channel accurately using real underwater data collected from a water tank with disturbance and from Lake Tahoe. We leverage the capability of reservoir computing for modeling dynamical systems and provided a data-driven approach to modeling the UWA channel using Echo State Network (ESN). In addition, the potential application of transfer learning to reservoir computing has been examined. Experimental results show that ESN is able to model chaotic UWA channels with better performance compared to popular deep learning models in terms of mean absolute percentage error (MAPE), specifically, ESN has outperformed deep neural network by 2\% and as much as 40\% in benign and chaotic UWA respectively.
\end{abstract}


\titlepgskip=-15pt

\maketitle

\section{Introduction}
\label{sec:introduction}
\PARstart{U}{nderwater} wireless communication has rapidly grown in importance for numerous ocean monitoring and information exchange applications in civil and military use in recent years~\cite{MAUACCB2017}. Acoustic technology has also been shown as a useful tool for a wide range of underwater activities and applications. Ocean exploration, scientific data collection, and transmission are some of the most prevalent applications for underwater acoustic (UWA) communications. Furthermore, underwater communications have benefited the maritime sector by making process management and monitoring easier and more efficient~\cite{UACHabors2019}.  Underwater operations including undersea marine biology study, undersea mining, pipeline laying, underwater maintenance, and geological surveys have fueled the increased demand for underwater channel and environment research~\cite{MACUWCS2012}. In general, the behavior of the channel has a significant impact on acoustic signal transmission, therefore having a deep understanding of the channel characteristics is critical for implementing an effective underwater communication system~\cite{UACHabors2019}.

\begin{figure}[h] 
	 \centering
    	 \includegraphics[width=\linewidth]{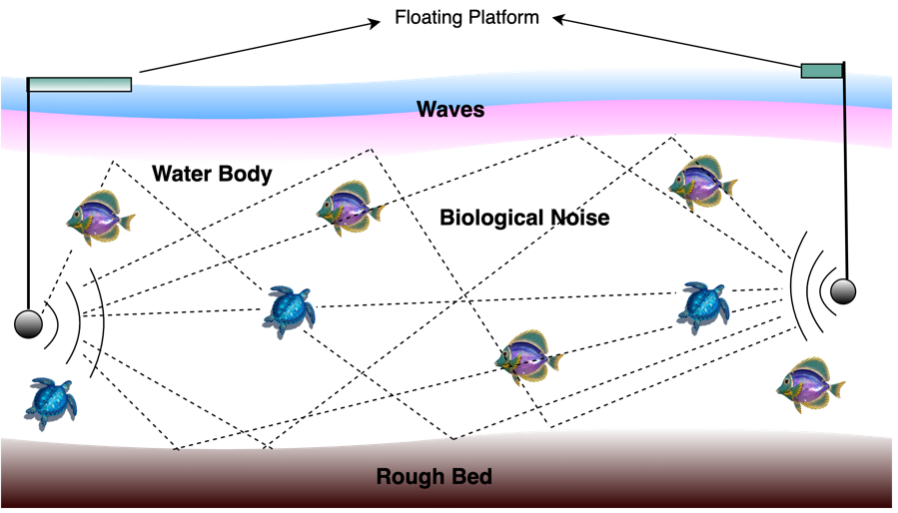}
      	\caption{A Typical Underwater Environment~\cite{LuoCognitive2014}}
    \label{fig:underwater_image}
\end{figure}

The underwater environment presents a unique set of challenges for wireless communications~\cite{MAPCUWSN2014}, and UWA channels are widely regarded as one of the most challenging communication media now in use~\cite{MAUACCB2017}. While low frequencies are excellent for acoustic propagation, the bandwidth available for communication is extremely limited. Furthermore, a UWA channel has low physical link quality and high latency, and it suffers from large multipath delay spread and frequency selective fading~\cite{chanmodelUAN2020}, making modeling of the UWA channel quite challenging~\cite{chanmodelIOT2021,UWAAlgorithm2019,MAPCUWSN2014,StojanovicUWA2009, chanmodelUWCN2008}. A typical UWA communication scenario is depicted in Figure~\ref{fig:underwater_image}. As illustrated in the figure, variations in sound velocity, roughness of the ocean bed, multi-path propagation of acoustic signals, and ambient ocean acoustic noises created by aquatic creatures and human activities make it even more challenging to model UWA channel~\cite{UACHabors2019,StojanovicUWA2009}. 

Many physics-based UWA channel models have been developed. The most commonly used one is the BELLHOP model, which is an open-source beam/ray-tracing model for predicting acoustic pressure fields in the underwater environment~\cite{StojanovicUWA2009}. 
The BELLHOP ray model is an intuitive and straightforward means for modeling sound propagation in the ocean among the various existing mathematical UWA channel models based on ray, normal-mode, and parabolic curve~\cite{MAUACCB2017}. 
The majority of these models, however, are based on mathematical assumptions and approximations rather than real underwater communication data. As a result, they do not work well in reality~\cite{MAUACCB2017}.

Machine learning has seen considerable success in fields such as image and voice recognition, language processing, medical diagnosis, and wireless communications. This is due largely to its capacity to learn and intelligently respond to changing and complex operating conditions, such as those found in the UWA communication channel. Specifically, it is shown that modeling the UWA channel by replicating the effect of real water environment characteristics on the channel is effective~\cite{MAUACCB2017}. However, there has not been much research work done in the domain of UWA communications using machine learning because of the complex nature of the underwater environment and the lack of sufficient and high-quality data. This motivated us to leverage the capability of collecting real-world data of UWA communications from Lake Tahoe in Reno, Nevada, and apply a data-driven approach to modeling the UWA channel using machine learning on the collected datasets~\cite{UACCMUDL2021}. 

RC is explored to model UWA channel in this paper. It has been shown that RC is capable of modeling dynamical systems~\cite{phoneme2010,infoProcessing2011} and predicting chaos~\cite{modelFree2018}.   
RC is a time-dependent data processing paradigm influenced by neuroscience~\cite{8261502}. It is a type of recurrent neural network model in which the recurrent component is initialized randomly and subsequently fixed thus incurring less computation and reducing training time~\cite{RCApproaches2009,Randomness2017}. Despite this significant simplification, the recurring element of the model, the reservoir has a huge number of dynamic properties that can be used to solve a wide range of problems~\cite{RCapproaches2018}. 

\begin{figure}[htbp]
    \centering
        \includegraphics[width=\linewidth]{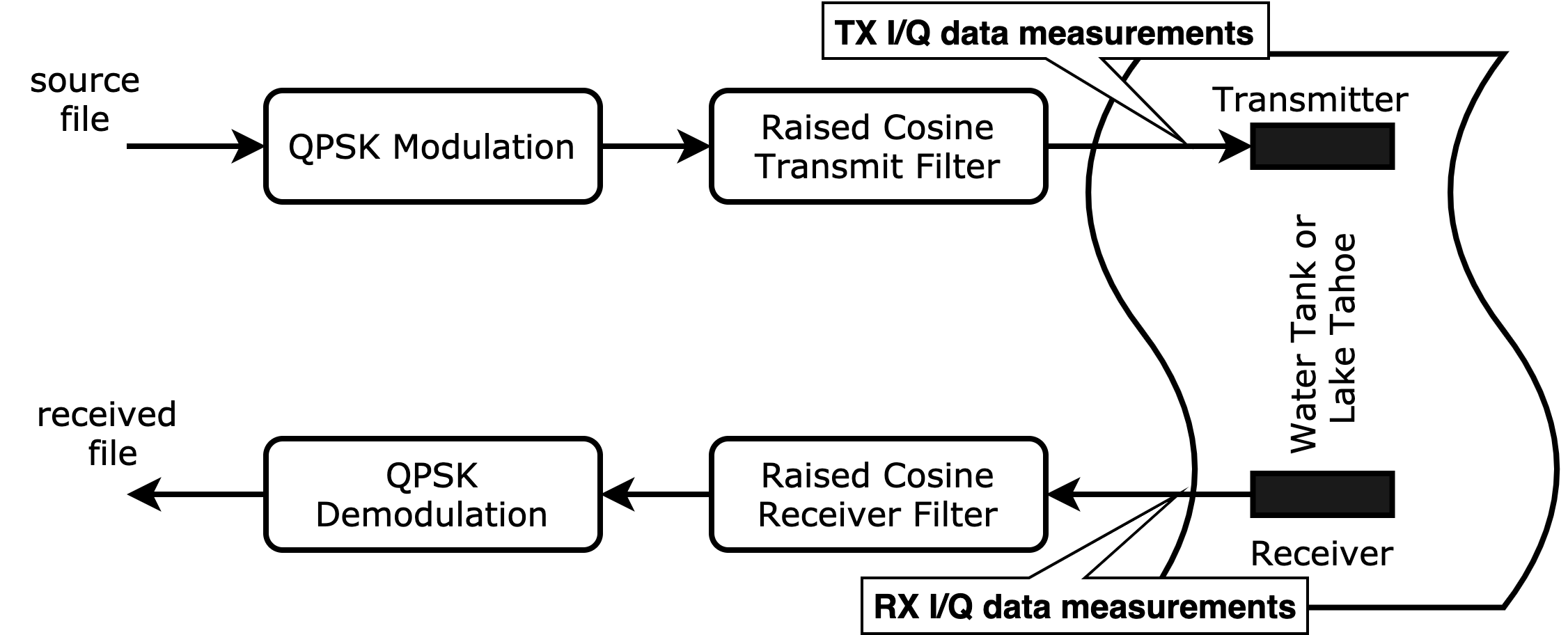} 
\caption{Block diagram of the UWA communications data collection.}
\label{fig:block}
\end{figure}

In this work, we seek to leverage these properties of RC and provide a data-driven approach to modeling the UWA communication channel using the Echo State Network (ESN), which is one of the two pioneering RC approaches. Using this approach significantly reduced the implementation complexity and the training time. The contributions of this paper are:
\begin{enumerate}
    \item A data-driven approach for UWA channel modeling is proposed to take advantage of the real-world experimental UWA datasets and avoid the unrealistic assumptions made by physics-based mathematical models.
    \item UWA channel modeling using ESN, an approach of RC and transfer learning have been carried out. Observations and insights are provided based on the experimental results. 
\item The effects of different setups of the ESN on the model performance, in terms of the reservoir initialization method, the size of the reservoir, the activation function, and the regression algorithms used at the readout layer, have been examined and suggestions are made to improve the performance of ESN for UWA channel modeling.
    \item A novel approach of designing the reservoir using a pre-trained deep learning model as the reservoir has been proposed in this study. Experimental results demonstrate that ESN using pre-trained deep learning models as reservoir outperform the deep learning models for modeling the UWA channel. Although this may not be advantageous if the reservoir of the ESN can be designed properly using randomized weights, it provides a systematic way to set up the reservoir, which is valuable because there does not exist a systematic way for the design of reservoir for diverse real-world applications. 
        \item Transfer learning using simulated radio frequency (RF) data and Bellhop-based simulated UWA channel data has been performed. It is observed that the performance of transfer learning using RF data is poor because of the significant differences in characteristics between RF channel and UWA channel. On the contrary, the performance of transfer learning using Bellhop-generated data is pretty good, although the Bellhop model itself is a simplified mathematical model and it does not take into account the various uncertainties in a real-world scenario.
\end{enumerate}

The remainder of this paper is structured as follows. Some related works are discussed in Section~\ref{sec:relatedwork}. Section~\ref{sec:datas} describes the data generation and collection process and gives the description of the datasets used in this work. Section~\ref{sec:reservoir} discusses RC and the ESN approach used in this experiment. Detailed experimental results and analysis are given in Section~\ref{sec:results} and observations and insights from the results are discussed. Section~\ref{sec:concl} concludes the paper.

\section{Related Works}
\label{sec:relatedwork}
Many mathematical models have been built and used for different purposes in theUWA domain including but not limited to investigating and tracking channel properties, sound propagation characteristics, the behavior of acoustic signals for different transmission frequencies, and the computation of some channel parameters such as the route loss. For example, in~\cite{MAUACCB2017}, a BELLHOP ray model was used to model the UWA channel in order to examine sound propagation characteristics while taking into account the rough nature of sea surfaces and bottoms for various oceanic conditions.
It was also used to examine the behavior of an acoustic signal with transmission frequencies in the range of 9K to 90KHz in~\cite{UACMUBRTM2017}.  
Channel properties for an autonomous underwater vehicle (AUV) wireless communication system were mathematically quantified by modeling the UWA channel in~\cite{MACUWCS2012}. The model was created using the $AN$ product, signal-to-noise ratio (SNR), and band selection, where $A$ represents attenuation and $N$ represents ambient noise.  In~\cite{chanmodelUWCN2008}, simulations using ray-theory-based multipath Rayleigh underwater channel models for shallow and deep waters are carried out to investigate transmission losses between transceivers, the effects of bit error rate, maximum internode distances for different networks and depths, the effect of weather season, and variability of ocean environmental factors.  The authors in~\cite{DUACTCRTVC2021} proposed a channel model for tracking dynamic UWA channels by using the channel's  correlation as the state-space model in the Kalman filter in order to improve tracking. The authors of~\cite{chanmodelIOT2021} calculated the channel route loss, changed the log-distance model to produce a model suitable for an underwater IoT network, and created an empirical channel model for medium-distance UWA channels based on real measurement data.  In~\cite{zhu20212d},  a non-stationary two-dimensional wideband channel model was designed for UWA communication and evaluated with measurement data. 

Various deep learning models have been proposed to model UWA communication channels. For example,
in~\cite{SDFCUACMDLM2019}, a deep learning network based signal detection was employed for full-duplex cognitive UWA communication with self interface cancellation. Automatic modulation classification of underwater communication signals using a combination of the convolutional neural network (CNN) and LSTM~\cite{ModClassUACDL2018}. A similar task has been done in~\cite{BEAMCUAS2018} using blind equalization in conjunction with a CNN. 
Because it is difficult to identify modulation during actual communication due to the complex and unstable nature of UWA communication systems, several machine learning methods were used in~\cite{LAUCNML2020} to classify the modulation type in their quest to find an efficient link adaptation method based on channel quality of an underwater communications network.
In~\cite{DLBSCCUAC2019}, the authors employed a DNN to create a deep learning-based receiver for single carrier communication in a UWA  channel utilizing data from the sea. When compared to the traditional channel-estimate based decision feedback equalizer, the DNN based receiver consistently performed better. The authors in~\cite{DNNChanEstUACOFDMRongkun2019} also utilized the DNN to estimate channel parameters based on data from the Bellhop Ray model simulation of the UWA environment. When compared to traditional channel estimation methods such as least square and minimum mean square error (MMSE), the DNN outperformed the LS algorithm and is comparable to the MMSE algorithm in terms of bit error rate and normalized mean square error.
A deep learning-based UWA orthogonal frequency-division multiplexing (OFDM) communication system was constructed by representing the receiver as a DNN in~\cite{ZHANG201953} and~\cite{UndAcouComDeepLearnYouwen2018}. The deep learning UWA communication systems could easily recover the transmitted symbols after training without using explicit channel estimation and equalization. 
 In~\cite{deepUnderWater2018}, the authors developed a depth learning-based underwater target recognition approach employing CNN and an extreme learning machine  for UWA target classification and recognition. 
 
RC has been applied in wireless communications for predicting wireless channel or state conditions,  symbol detection, and measuring the channel SNR. For example, in~\cite{9002360}, the performance of an extreme learning machine and an ESN for forecasting wireless channel conditions was compared. These two methods were used to forecast the SNR for single-input single-output systems in both pico-cellular and micro-cellular contexts.  For multiple-input multiple-output orthogonal frequency-division multiplexing (MIMO-OFDM) systems, an ESN-based symbol detector was used in~\cite{Braininspired2017}. The efficiency of the adopted symbol detector outperforms traditional symbol detectors based on channel estimation methods in terms of BER performance according to simulation results. 
In~\cite{9020011}, a new RC-based detector called windowed ESN was designed for MIMO-OFDM symbol detection. This resulted in significant improvements in interference cancellation and nonlinear compensation, as well as the ability to improve short-term memory fundamentally. 
The authors in~\cite{8261502}  looked at a simplified fading channel model, defined the transmission properties of satellite communication channels, and devised an ESN-based approach for measuring channel SNR. For the categorization of multivariate time series, the authors in~\cite{bianchi2020reservoir} applied an unsupervised approach for creating multivariate time series (MTS) representations (also known as reservoir model space). The parameters of a one-step forward predictor that forecasts the future reservoir state rather than the future MTS input were used to create the reservoir model space. The results revealed that RC classifiers are substantially faster and achieve higher classification accuracy. In~\cite{ChanEstim2015}, an ESN was utilized to train an RNN to predict channel state information in a wireless OFDM system, which resulted in a significant decrease in training time, implementation, and computing complexity. 
RC has also been applied in UWA communications. For example, transfer learning was introduced to ESNs in~\cite{chen2021predicting} to develop a channel model that predicts shallow water dynamics. Experimental results showed that transfer learning helped improve the predictions.  

In this work, we leveraged the capabilities of RC and deep learning to build a data-driven channel model using a real experimental UWA communications dataset collected from Lake Tahoe under various environmental conditions. Different from the existing works, the goal of this work is to model the UWA channel and be able to perform sequence-to-sequence prediction, i.e.,  when a sequence of transmitted data is fed into the model as input, the corresponding sequence of data is expected to be received at the receiving end of the UWA channel will be predicted. The obtained UWA channel model in this work would be very useful as a candidate plugin module when large scale simulations of UWA communications are needed, or a large amount of data need to be generated for UWA channels with high fidelity while it is difficult to obtain that kind of data from physical underwater data collections.

\section{UWA Dataset}
\label{sec:datas}

\subsection{Real-World UWA Communications Data Collection}
\label{sub:datagen}
\begin{figure}[ht]
	 \centering
    	 \includegraphics[width=\linewidth]{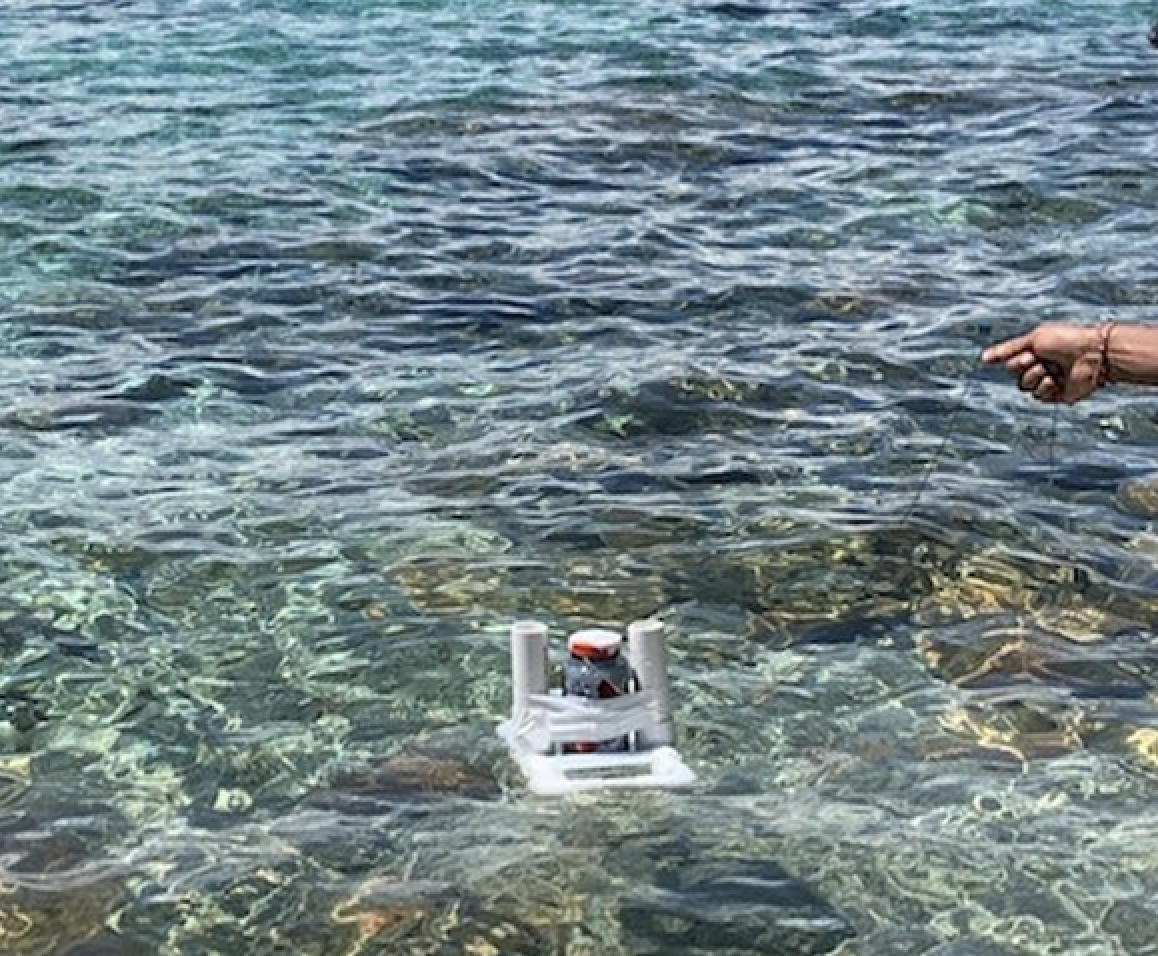}
      	\caption{The test bed at Lake Tahoe.}
    \label{fig:water}
\end{figure}

Underwater communication testbeds were built to collect the UWA communications data for training and further evaluate the developed learning-based channel modeling. To fully study the developed technique, a series of experiments have been conducted. It includes the lab-based experiment and the open-water test and the set-ups for these experiments are as summarized in the block diagrams in Figure~\ref{fig:block}. In the lab-based experiment, a water tank was used. For the open-water experiment, the experiments were carried out at Lake Tahoe. Lake Tahoe, as seen in Figure~\ref{fig:water}, is a large freshwater lake in the Sierra Nevada Mountains that straddles the California-Nevada state line. According to Wikipedia, it is the largest alpine lake in North America and at a maximum depth of $1,645$ feet (501 meters), it is the second deepest lake in the United States. Lake Tahoe is also said to be the $16$th deepest lake in the world, and the fifth deepest in average depth. This work does not target sea or ocean environment which has its unique characteristics such as the ocean salinity and we recognize that some modifications and refining the model might be needed to properly transfer it for the ocean environment. By taking advantage of our access to Lake Tahoe, a very large lake that has a lot of similar characteristics as the undersea environment, such as waves, aquatic-life disturbances, etc. the developed model may be served as a reference model to be modified and transferred to ocean environment. In the future, authors plan to further evaluate the developed algorithm in more uncertain environments such as open sea.

\begin{figure}[h]
	 \centering
    	 \includegraphics[width=\linewidth]{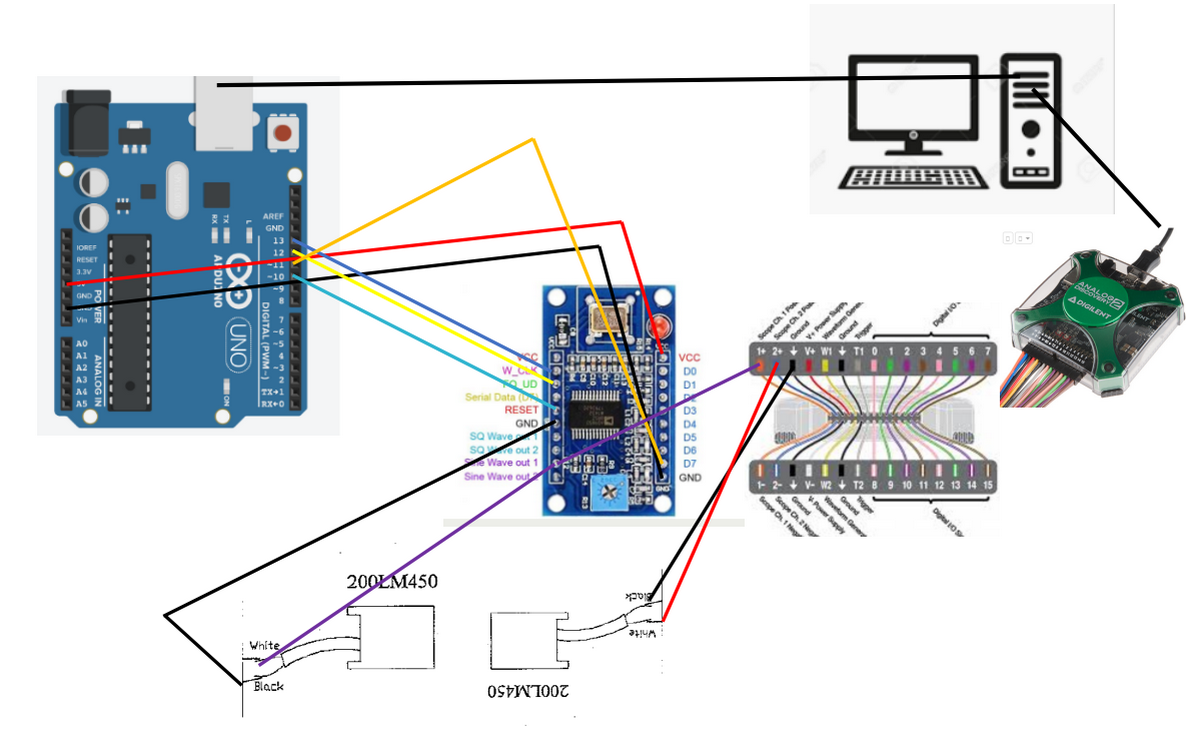}
      	\caption{Transmitter Set-Up} 
        \label{fig:trans}
    	 \includegraphics[width=\linewidth]{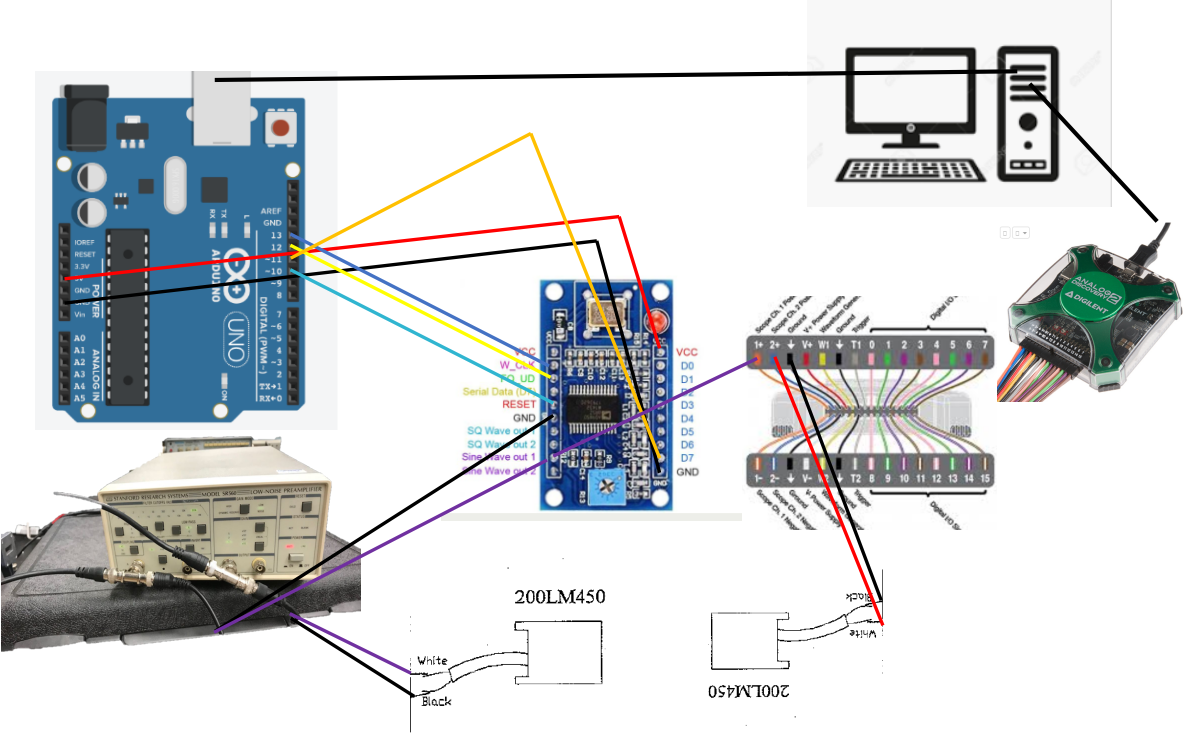}
      	\caption{Receiver Set-Up}
    \label{fig:rec}
\end{figure}

For the developed underwater data acquisition system, the transducer is made watertight and other electronic systems including MCU have been packed inside a watertight enclosure tube from BlueRobotics Inc. , the manufacturer. To collect enough underwater communication data, we have deployed transducers at 1-2 meters below the water surface and the distance between Tx and Rx is around 3-5 meters. For each experiment scenario, the source signals were first coded through an oscillator along with the microcontroller (MCU) module. The coded signal strength was then enhanced through an amplifier circuit. It was important to strengthen the signals because the strength of the transmitted signals changes or drops along with the communication distance underwater. To stabilize the signal strength along the transmission path, the amplifier limiter circuit has been used to amplify or enhance the signal strength. Next, the enhanced signals were passed through a quadrature phase shift keying (QPSK) modulation block which outputs continuous signals. The continuous signals were then passed through a raised cosine transmit filter and finally to an ultrasonic ceramic transducer (200LM450) which doubles as the transmitter, as shown in Figure~\ref{fig:trans}. The transmitted I/Q data has been collected at the transmitter and used as input during the training of the channel model. The transmitter and the receiver were placed horizontally apart and at a perpendicular distance below the water level. 

At the receiver end, the transmitted continuous signals were captured by the receiver/ ultrasonic ceramic transducers (200LM450). The received I/Q data has been collected at the receiver and used as ground truth for the training of the channel model.  The received signals are then reformulated, demodulated, and decoded through the automatic gain control (AGC) circuit as well as a series of filters including the raised cosine receive filter. Using the Schmitt trigger with an analog to digital converter (ADC) circuit, the received signals can be digitalized for further decoding in the workstation. The collected time-series data at the receiver corresponds to the channel impulse response (CIR) of the UWA channel.The receiver setup is as shown in Figure~\ref{fig:rec}. The sampling rate of acquisition used is $1,000,000$ and the length of each data object is $60$ seconds. The sonar working frequency is $200$kHz and the digital signal transfer speed is $2K/s$.  Research has shown that deep learning requires large amount of data to obtain good performance~\cite{DeepLargeData2020}.  To collect sufficient data to train our deep learning models and for our models to work well,  we used a sampling rate of 1 MHz to obtain time-series data containing 60 seconds of I/Q samples.

\subsection{Data Characteristics}
For the purpose of modeling the UWA channel using machine learning and RC models, data collected at the transmitter (ultrasonic ceramic transducer) just before the channel were used as the input to our models while the data collected at the receiver, immediately after the channel were used as the ground truth for training the models. Four different data categories were collected and used to train, evaluate and compare the performances of the trained models. The datasets are described as follows: The first category of data, subsequently referred to as Data 1, were collected using the water tank as the communication channel with no external disturbance.  The second category of data collected termed Data 2 were also collected from the lake with no artificial/external disturbance introduced. The third category of data termed Data 3 were also collected from the lake but with the introduction of mild external disturbance. 

The disturbance was introduced to create a more realistic underwater scenario. The fourth category, termed Data 4, was also collected from the lake but with the introduction of strong external disturbance to mimic a more chaotic underwater scenario.  To model the waves or disturbance in the lake, a vibration platform was used to inject the vibration that generated waves in the lake. For all the categories of data, $60,000,000$ samples were collected using the same transmission settings and parameters. These data descriptions are summarized in Table~\ref{tab:description} below.

\begin{table} [h] 
\caption{UWA Data Description}
\label{tab:description}
\centering
\begin{tabular}{llccc}
\toprule
\textbf{Category}&\textbf{Channel}&\textbf{External Disturbance}&\textbf{Size}\\
\midrule
Data 1 & Water Tank & None & 100,000$\times$578 \\ 
Data 2 & Lake & None & 100,000$\times$578 \\ 
Data 3 &Lake & Mild & 100,000$\times$578 \\ 
Data 4 & Lake & Strong & 100,000$\times$578 \\ 
\bottomrule
\end{tabular}
\end{table}
 
\section{Reservoir Computing}
\label{sec:reservoir}
RC is a time-dependent data processing paradigm influenced by neuroscience~\cite{8261502}. It is a class of recurrent neural network (RNN) model in which the recurrent component is produced randomly and subsequently fixed~\cite{RCApproaches2009,Randomness2017}. The RC methodology builds an RNN with random synaptic weights, dubbed the reservoir, in order to avoid the gradient-descent procedures of the training algorithms for a typical RNN. In~\cite{RCApproaches2009} and~\cite{chaoticneuralnet2009}, RC shows how an RNN with fixed connectivity can memorize and produce complicated Spatio-temporal sequences. RC has also been demonstrated to be a valuable tool for modeling and predicting dynamic systems~\cite{phoneme2010,infoProcessing2011}. It was demonstrated in~\cite{modelFree2018} that RC is capable of forecasting massive chaotic systems. 

Despite this significant simplification, the recurring element of the model (the reservoir) has a huge number of dynamic properties that can be used to solve a wide range of problems~\cite{RCapproaches2018}. It is also pertinent to state that RC involves less computation and thus has reduced training time dramatically.  The normal workflow of solving a task using RC requires handling two key steps: (1) designing a suitable reservoir for the specific task under consideration, and (2) determining a readout function that will adequately map the state of the reservoir to a target output~\cite{Singer2021}. One of the concerns of RC is that the design is mainly driven by a succession of randomized model-building stages, leaving researchers to rely on a series of trials and errors~\cite{5629375}. In this work, a novel approach to designing the reservoir using a pre-trained deep learning model as the reservoir has been proposed. Although this may not be advantageous if the reservoir of the ESN can be designed properly using randomized weights, it provides a systematic way to set up the reservoir, which is valuable because there does not exist a systematic way for the design of reservoir for diverse real-world applications. 

There are two popular RC approaches: the Liquid State Machines (LSM) and ESN~\cite{kleyko2020integer,RTCWSS2002,ANSIWESN2003}. Both architectures attempt to model biological information processing using similar principles~\cite{analysis2007}. In this work, ESN is chosen because of its close relationship to RNN/LSTM and allows direct performance comparison between these different models.

\subsection{Echo State Network}
\label{sec:echo}
\begin{figure}[htbp]
	 \centering
    	 \includegraphics[width=\linewidth]{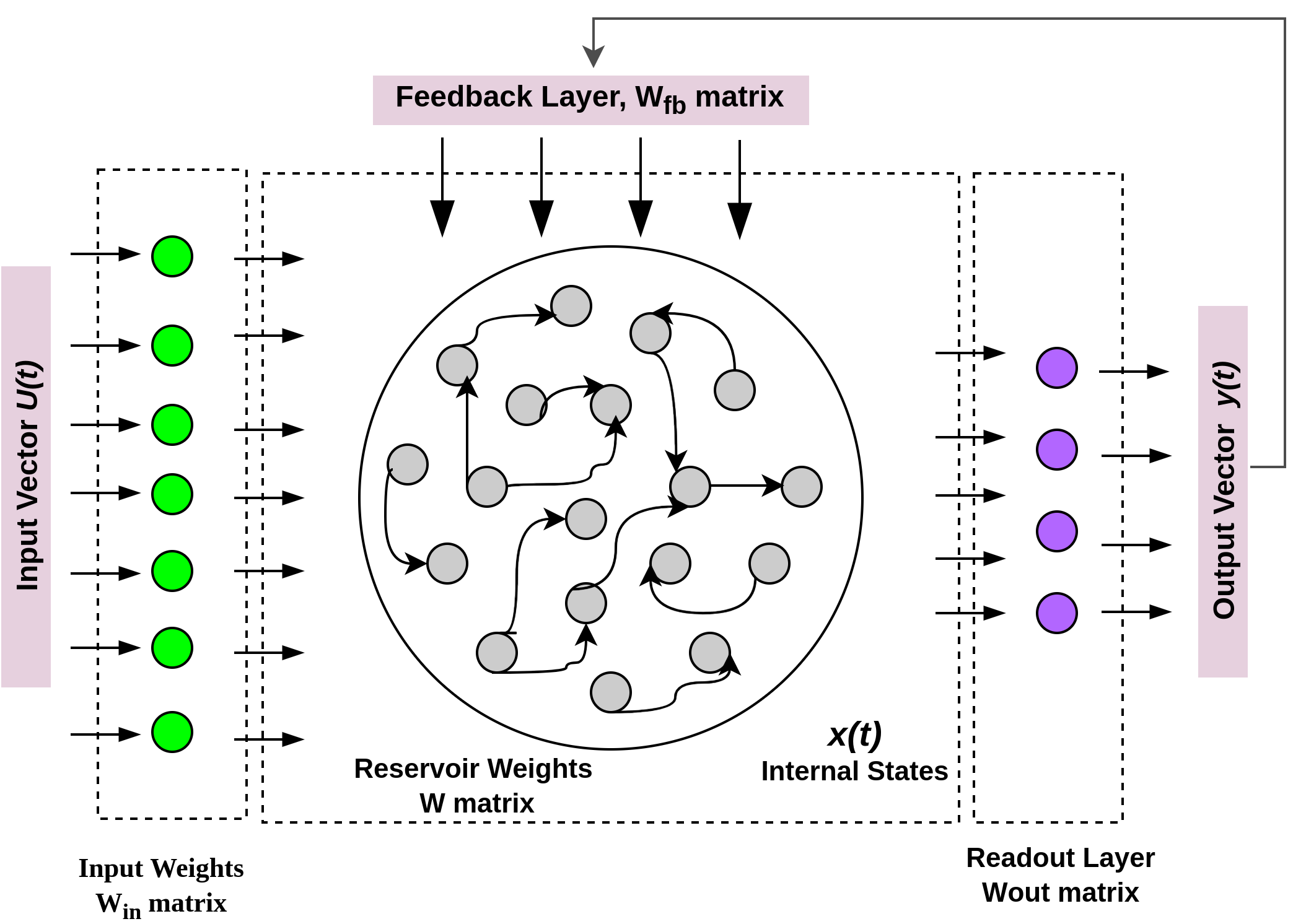}
      	\caption{Typical architecture of an ESN~\cite{ReservoirPy2020}}
    \label{fig:esn}
\end{figure}

ESNs are dynamical artificial neural networks and belong to the general class of RNN. This approach is prominent and is based on the discovery that if a random RNN has certain algebraic features, then training a linear readout from it is typically enough to provide outstanding performance in practical applications~\cite{Cernansk2008PredictiveMW}. ESNs have a topology of nonlinear processing elements that is densely interconnected and recurrent, forming a ``reservoir'' that stores information about the history of input and output patterns. The outputs of these internal processing elements are referred to as the ``echo states''. The titles of the echo states stem from the input values echoing throughout the reservoir's states due to the reservoir's recurrent nature~\cite{RCApproaches2009, RC2014}. These echo states are fed into a memoryless but adaptive, usually linear, readout network, which generates the network output. The architecture of a typical ESN is shown in Figure~\ref{fig:esn}. ESN has the unique property of just training the memoryless readout, whereas the recurrent topology has fixed connection weights. This reduces the complexity of RNN training to simple regression while maintaining the recurrent topology, but at the same time, it imposes significant constraints on the overall architecture that has yet to be resolved~\cite{analysis2007}. 

As shown in Figure~\ref{fig:esn}, ESN has three layers; the input layer, the dynamic reservoir, and the output layer. The input weight matrix $W_{in}$ connects the input layer to the dynamic reservoir. The internal weights of the dynamic reservoir, $W$  define the linkages inside the reservoir. The output weight matrix $W_{out}$ connects the dynamic reservoir to the output layer. Feedback weights $W_{fb}$ are used to feed the output back into the dynamic reservoir. The fundamental structural distinction between an ESN and the conventional RNN is the connectivity of neurons within the dynamic reservoir~\cite{Braininspired2017} and one major advantage of RC over regular RNN is that simple regression algorithms may be used to alter output weights~\cite{Cernansk2008PredictiveMW}.

Through the weighted input connections, the input layer of neurons delivers the stimulus to stimulate the reservoir. Through the weighted feedback connections, the output layer of neurons transmits teacher-forced outputs to the reservoir. The reservoir is trained to generate weighted connections from the reservoir to the output based on the input stimuli and feedback from the teacher-forced outputs~\cite{ChanEstim2015}. Specific values are used to weigh the connections between each layer of neurons and between neurons in the reservoir. Let us consider a recurrent discrete-time network with $K$ input units, $N$ internal processing elements, also known as nodes and $L$ output units. The value of the input unit at time, $t$ is $\mathbf{U} = \mathbf{[}\mathbf{u}_1(t), \mathbf{u}_2(t), ..., \mathbf{u}_K(t) \mathbf{]}$, the value of the internal units is $\mathbf{X} = \mathbf{[}\mathbf{x}_1(t), \mathbf{x}_2(t), ..., \mathbf{x}_N(t) \mathbf{]}$ and those of the output units are $\mathbf{Y} = \mathbf{[}\mathbf{y}_1(t), \mathbf{y}_2(t), ..., \mathbf{y}_L(t) \mathbf{]}$. An $N \times K$ matrix, $\mathbf{W}_{in}$, defines the weights of connections from the input layer to the dynamic reservoir $\mathbf{W}$, which is an $N\times N$ matrix for connection between the nodes. Also, an $L \times N$ matrix,  $\mathbf{W}_{out}$ defines the connection from the reservoir nodes or processing elements to the output units. Lastly, $\mathbf{W}_{fb}$,  which is  an $N\times L$ matrix defines the connection weights of the feedback from the output layer to the reservoir~\cite{analysis2007}. Only the output weights $\mathbf{W}_{out}$, are computed during training, while the rest of the connection weights are generated randomly and fixed throughout the training and testing stages\cite{ChanEstim2015}. In Figure~\ref{fig:esn}, assuming $\mathbf{u}(t)$ is the input vector at time step $t$, the activations of hidden nodes, also known as the echo states, $\mathbf{x}(t)$ are updated according to equation~(\ref{eqn:three}).
\begin{equation}
\mathbf{x}(t) = \mathbf{f}(\mathbf{W}_{in} \mathbf{u}(t) + \mathbf{W}\mathbf{x}(t-1) +\mathbf{W}_{fb} \mathbf{y}(t-1))
   \label{eqn:three} 
\end{equation} 
where $\mathbf{f}$ is an hyperbolic tangent activation function of the hidden unit, $\mathbf{W}$, $\mathbf{W}_{in}$  and $\mathbf{W}_{fb}$  are the matrices of hidden-hidden, input-hidden, and output-hidden connections, respectively~\cite{analysis2007}. Equation~(\ref{eqn:three}) can thus be rewritten as
\begin{equation}
\mathbf{x}(t) = \mathbf{tanh}(\mathbf{W}_{in} \mathbf{u}(t) + \mathbf{W}\mathbf{x}(t-1) +\mathbf{W}_{fb} \mathbf{y}(t-1)) \;
 \label{eqn:four} 
\end{equation}
This equation is also known as the state transition equation.

\subsection{Training the Readout Layer}
\label{readout}
As mentioned in Section~\ref{sec:echo}, when training the ESN, only the output connection matrix $\mathbf{W}_{out}$ is updated while other connection matrices are kept constant. The training process involves driving the reservoir with the input time series data, $\mathbf{u}(t)$ to generate the corresponding states, $\mathbf{x}(t)$ using Equation~(\ref{eqn:four}). All the states are then collected into a matrix $\mathbf{X}$ and the target data (ground truth) are collected into another matrix $\mathbf{\hat{Y}}$. The output weights are then computed in closed form using Equation~(\ref{eqn:wout}) given by
\begin{equation}
\mathbf{W}_{out} = \mathbf{\hat{Y}}\mathbf{X}^T(\mathbf{X}\mathbf{X}^T + \mathbf{\lambda}\mathbf{I})^{-1}
   \label{eqn:wout} 
\end{equation} 
where $\mathbf{I}$ is an identity matrix and $\mathbf{\lambda}$ is the \textbf{Tikhonov} regularizer which is a fixed positive number used to determine the sensitivity of the system~\cite{AMFESN2011}. Since we are dealing with batches of sequence data, Equation~(\ref{eqn:wout}) becomes
\begin{equation}
\mathbf{W}_{out} = \mathbf{A}(\mathbf{B} + \mathbf{ \lambda} \mathbf{I})^{-1} \\
   \label{eqn:batches} 
\end{equation} 
where   $\mathbf{A} = \sum_{i}^{} \mathbf{\hat{Y}}_i\mathbf{X}_i^T$ and $\mathbf{B}  = \sum_{i}^{} \mathbf{X}_i\mathbf{X}_i^T$. The output from the readout layer is also computed using the simple output layer equation given by
\begin{equation}
 \mathbf{y}(t) = \mathbf{f}_Y(\mathbf{W}_{out} \cdot \mathbf{x}(t) )
   \label{eqn:five} 
\end{equation} 
where $\mathbf{f}_Y$ is the output nonlinear activation function~\cite{Cernansk2008PredictiveMW}. The task of training the readout is then reduced to a simple linear regression problem of minimizing the squared error.  The regression model minimizes the mean square error between predictions, $\mathbf{Y}$ and the ground truth, $\mathbf{\hat{Y}}$, i.e., $ \left\| \mathbf{Y}  - \mathbf{\hat{Y}}\right\|_2^2$ .

\subsection{Echo State Property}
\label{echoppty}
In RC, one important criterion that must be satisfied is the echo state condition, especially when working with ESN. In essence, this condition states that the effect of a prior state and a previous input on a future state should fade away or vanish with time. In other words, it is stated that the dynamics of the ESN should be uniquely controlled by the input~\cite{analysis2007}. The echo state condition is defined in terms of the spectral radius, $\rho(\mathbf{W})$ of the reservoir weight matrix, $\mathbf{W}$. 
Specifically, 
assuming $\mathbf{\lambda}_1,\mathbf{\lambda}_2,...,\mathbf{\lambda}_n$ are the eigenvalues of the reservoir matrix $\mathbf{W}$, then its spectral radius, $\rho(\mathbf{W})$ is defined as
\begin{equation}
\rho(\mathbf{W}) = \left\| \mathbf{W}\right\|  =   \max {\begin{Bmatrix} \begin{vmatrix}\mathbf{\lambda}_1\end{vmatrix},\begin{vmatrix}\mathbf{\lambda}_2 \end{vmatrix} ,...,\begin{vmatrix}\mathbf{\lambda}_n \end{vmatrix}  \end{Bmatrix} }  \; .
\label{eqn:esp}
\end{equation}
The echo state condition is satisfied if $\mathbf{W}$ is scaled such that its spectral radius $\rho(\mathbf{W})$ is close to or inferior to $1$ as expressed in Equation~(\ref{eqn:esp}) given by
\begin{equation}
\rho(\mathbf{W}): \left\| \mathbf{W}\right\| < 1 .
\label{eqn:esp}
\end{equation}

The spectral radius is the largest absolute eigenvalue of the matrix $\mathbf{W}$ and is a crude way of measuring how much memory the reservoir can hold, with small values indicating a short memory and large values indicating a longer memory, up to the point of over-amplification, when the echo state condition no longer holds~\cite{LIRWMESN2013,ReservoirPy2020}.
Another important consideration with the echo state property is that it must guarantee that the memory capacity is not reduced to zero since one of the advantages of using RNNs is their capacity to have a memory of the inputs. The reservoir output should be able to recreate the input with a $K$ steps delay~\cite{RC2014}. Spectral radius of W can be specified by a user and used backward in the design or initialization of the reservoir so that we could guarantee the performance. It is a design choice, and it does not depend on data. 
\begin{figure}[h]
	 \centering
    	 \includegraphics[width=\linewidth]{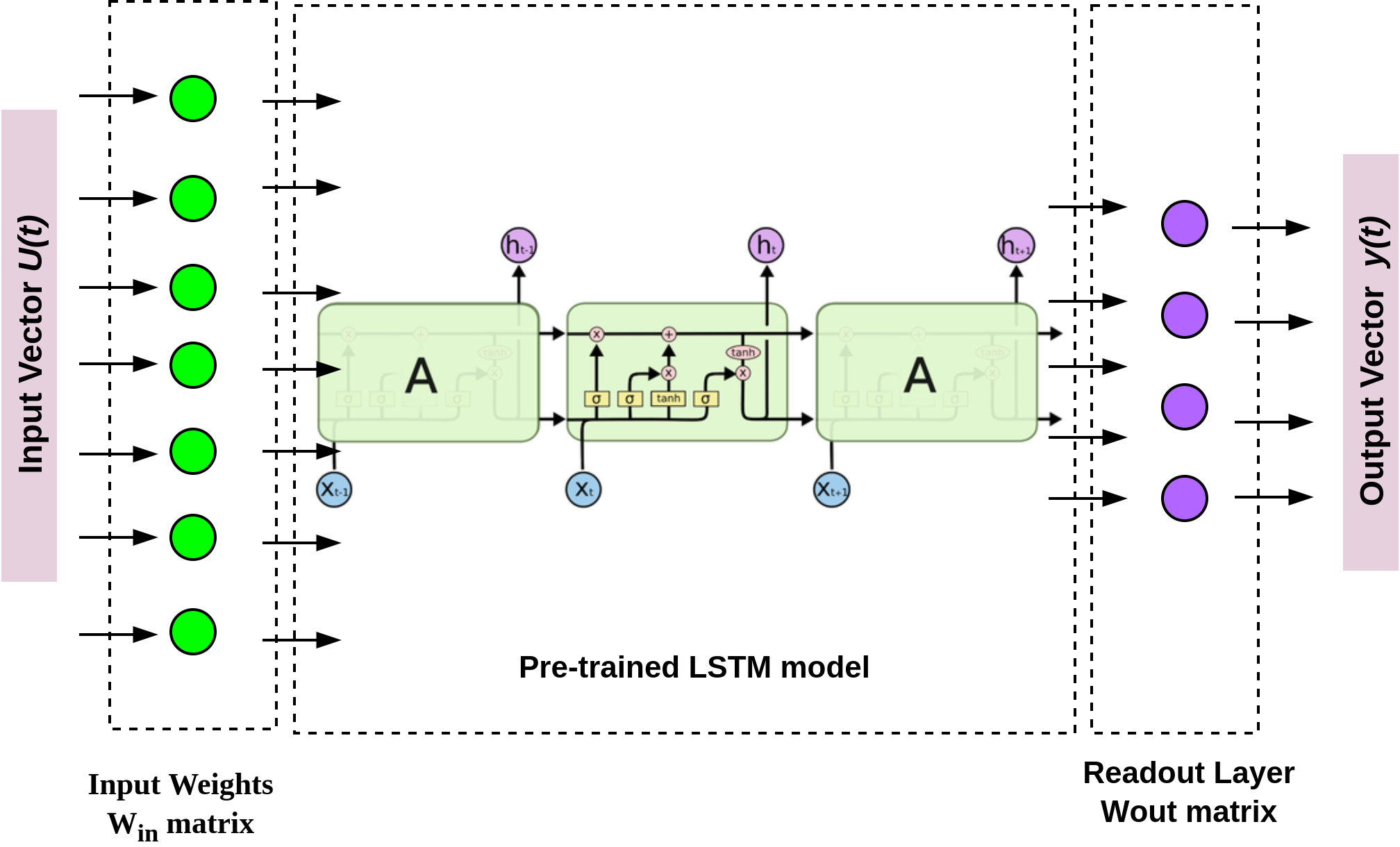}
      	\caption{An example ESN architecture with pre-trained LSTM as the reservoir.}
    \label{fig:esnlstm}
\end{figure}

One of the foci of this paper is using ESN for modeling the UWA communication channel. A known challenge with working with ESN is that there is no predefined systemic way of designing the dynamic reservoir for a specific application or use case. As of today, the design of the ESN relies heavily on the selection of the spectral radius. A suggested method of producing an appropriate reservoir, according to~\cite{Singer2021}, is to optimize their dynamics for the range of activities to be expected, such that the readout layer may simply extract the information it requires. The dynamic reservoir, $\mathbf{W}$ alongside $\mathbf{W}_{in}$ and $\mathbf{W}_{fb}$ are usually randomly generated at network initialization~\cite{Cernansk2008PredictiveMW},~\cite{kleyko2020integer} and stay fixed or left untrained during the network's lifetime~\cite{bianchi2020reservoir}. However, there are many different weight matrices with the same spectral radius that can be generated, and they don't all have the same performance with respect to the mean square error (MSE) or mean absolute percentage error (MAPE) for functional approximation. According to~\cite{analysis2007}, different randomizations with the same spectral radius perform differently on the same problem. To improve the performance of ESNs, many simple methods have been proposed. Some of these methods include increasing non-linearity by augmenting non-linear expansion with polynomial functions of reservoir activities, increasing reservoir size, averaging predictions from many reservoirs, introducing delay lines into the readout system, providing neurons with a diversity of time constants, and having the reservoir adapt to input statistics via intrinsic  plasticity~\cite{Laan2015EchoSN}. Also, the behavior of the reservoir, according to~\cite{RCapproaches2018} can be controlled by modifying the spectral radius $\rho(\mathbf{W})$, the sparsity (or the probability of connection), which is the percentage of non-zero connections, and the number of hidden units in the reservoir, $N$.

In this study, we investigated the effect of different initialization approaches for the reservoir with respect to the spectral radius, sparsity, and the number of hidden units and then compared their performances in terms of the MAPE. We also propose the use of  pre-trained deep learning models e.g. LSTM,  DNN, etc. into the ESN architecture to act as the reservoir instead of using some randomized vectors as the dynamic reservoir, thus leaving us with a modified ESN architecture such as one shown in Figure~\ref{fig:esnlstm}. The pre-trained deep learning models were trained with the same datasets. This architecture makes use of the weights from the pre-trained model and as with standard architecture of the ESN, only the output connections are modified during the learning process i.e training only occurs in the readout layer. The two established deep learning models experimented with are the deep neural network (DNN) and the long short term memory (LSTM).  The LSTM is a variant of RNN with the ability to learn input data with long-term dependencies~\cite{LSTM1997} while the DNN is a deep learning architecture with more than one hidden layer that are fully connected~\cite{BengioNOW2009}.

\section{Transfer Learning}
\label{sec:tla}
Transfer learning can be defined as the process of improving learning in a new task or domain by transferring knowledge acquired from another related task or domain that has already been trained. In other words, transfer learning relates to situations in which what has been learned in one domain is used to improve generalization in another domain~\cite{trcfafp2019}. 
In transfer learning, a base model is first trained on a base dataset and task. The learned features are then transferred to the target domain to be trained on a target dataset and task. This method is more likely to succeed if the features are generic, i.e., applicable to both the base and target domains, rather than being specific to the base domain. This approach is known as the ``pre-trained model approach''. The purpose of transfer learning is to quickly obtain the learning model by leveraging commonalities between tasks. In this study, transfer learning for RC has been considered. Specifically, transfer learning using simulated RF data and Bellhop based simulated UWA channel data has been performed. In this case, the base domain is RF wireless communications or simulated UWA communications, and the target domain is real-world UWA  communications.

\section{Results and Discussion}
In this section, we provide the experimental set up such as the hyperparameters' settings in section VI.A. Then the detailed performance of ESN is evaluated in section VI.B. Performance evaluations of modified ESN and transfer learning are carried out in section VI.C and section VI.D, respectively. Together they provide a comprehensive performance analysis of the proposed schemes.
\label{sec:results}
\subsection{Computing Experimental Setup}
Multiple sets of experiments were carried out and model performances are evaluated using the MAPE. The MAPE value is the average of all the absolute percentage errors in predictions. To compute the MAPE, percentage errors are added together without respect to sign, as shown in equation (\ref{eqn:seven}) where $A_t$ is the actual value and $F_t$ is the predicted value. 
\begin{equation}
   MAPE = \frac{\mathrm{1}}{n}  \sum_{t=1}^{n} \left| \frac{{A_t - F_t}}{A_t} \right|
   \label{eqn:seven}
\end{equation} 
MAPE provides a fairly intuitive interpretation in terms of relative error when evaluating regression problems, and it is preferable in the assessment since it provides the error in terms of percentages, avoiding the problem of positive and negative errors canceling each other out. The better the prediction, the smaller the MAPE and values recorded in the result tables below are the average value from multiple repeated experiments.  

\begin{table}[h]
\caption{Hyper-parameters for DL models and ESN}
\label{tab:hyperparam}
\centering
\begin{tabular}{clllll}
\toprule
\textbf{Table No.} & \multicolumn{1}{c}{\textbf{DNN}} & \multicolumn{1}{c}{\textbf{LSTM}}& \multicolumn{1}{c}{\textbf{ESN}} & \multicolumn{1}{c}{\textbf{PMESN}} \\ 
\midrule
 & $\alpha =  0.001$ & $\alpha =  0.001$&$I_M =  X$&\\ 
\ref{tab:deepvsecho},  \ref{tab:deepecho} & $H_{L}$ = 6 & $H_{L}$ = 6 & $N = 578$& \\
\ref{tab:rfdata},  \ref{tab:finetunedrf}& $L_{N} = 256$& $L_{N} = 256 $&$R_M = R_i$&$R_M = R_i$ \\
\ref{tab:belldata},  \ref{tab:finetunedbellhop}& $B_{S} = 64 $& $B_{S} = 64 $ & $A_F = H_T$& \\
& $N_{E} = 100 $& $N_{E} = 100 $&$\rho(W) = 0.5$& \\
\bottomrule
\end{tabular}
\end{table}

We started the set of experiments by training popular deep learning models such as DNN and LSTM and compare their performance with a typical setup of an ESN. The experimental results are given in Tables \ref{tab:deepvsecho},  \ref{tab:deepecho},  \ref{tab:rfdata},  \ref{tab:finetunedrf}, \ref{tab:belldata},  \ref{tab:finetunedbellhop}.
The hyper-parameters used for each of the models built in these experiments are summarized in Table~\ref{tab:hyperparam}, where the first column gives the index of the tables that contain the experimental results using these hyper-parameters. Here $\alpha$ is the learning rate,  $H_{L}$ is the number of hidden layers,  $L_{N}$ is the number of nodes per hidden layer,  $B_S$ is the batch size and $N_E$ is the number of epoch used.  

In order to study the effects of various hyper-parameters on ESN performance, a set of in-depth ESN experiments have been done. The experimental results are given in Tables \ref{tab:normradius},  \ref{tab:spectral}, \ref{tab:size}, \ref{tab:activation}, \ref{tab_regr}.
The hyper-parameters used for the different setups of the ESNs are listed in Table~\ref{tab:esnparam}. The first row gives the index of the tables that contain the experimental results using these hyper-parameters for ESNs. In the first column, $I_{M}$ is the initialization method used,  $\rho(W)$ is the spectral radius, $N$ is the reservoir size,  $A_F$ is the activation function used, and $R_M$ is the regression method used for training the output layer. 
$X$ is Xavier initialization method,  $G$ is normalized Xavier initialization (gloriot) method,  $HE$ is HE initialization method,    $Ri$ is ridge regression, $Li$ is linear regression, $La$ is lasso regression and $H_T$ is hyperbolic tangent. 
\begin{table}[h]
\caption{Hyper-parameters for in-depth ESN experiments }
\label{tab:esnparam}
\centering
\begin{tabular}{|p{1.4cm}||p{0.8cm}|p{0.8cm}|p{0.8cm}|p{0.8cm}|p{0.8cm}|}
\toprule
\multicolumn{6}{c}{\textbf{Table Number}}\\ \midrule
{\tiny \textbf{HyperParameters}}  & {\ref{tab:normradius}} & {\ref{tab:spectral}} & {\ref{tab:size}} & {\ref{tab:activation}} & {\ref{tab_regr}} \\ \midrule
$I_M$ & multiple & $X$ & $X$ & $X$ & $X$ \\ \hline
$\rho(W) $& $ 0.5$& multiple &$ 0.5$&$0.5$&$0.5$ \\ \hline
$N$ & $578$ & $578$ & multiple & $578$&$578$ \\ \hline
$A_F$& $H_T$& $H_T$& $H_T$& multiple &$H_T$ \\ \hline
$R_M $ & $R_i$&$R_i$&$R_i$ & $R_i$ & multiple \\ \bottomrule
\end{tabular}
\end{table}

\subsection{Performance Evaluation of the ESN}
\label{subsec:PerformanceEvaluationESN}
Two major categories of experiments, with several sub-experiments,  were carried out here to build and evaluate ESN models on different data quality and with different network setups. These experiments include
(1) comparison of performance between DL models and ESN; and (2) in-depth performance evaluation of ESN with different setups.

\subsubsection{Comparison of performance between DL models and ESN} 
Deep learning models and ESNs were built, trained, and evaluated using Data 3 (data from experiments in Lake Tahoe with mild disturbance) and Data 4 (data from experiments in Lake Tahoe with strong disturbance, hence has the poorest quality).  Model parameters used are provided in Table~\ref{tab:hyperparam}. Results from these experiments were recorded in Table~\ref{tab:deepvsecho}. It is observed that ESN outperforms DNN and LSTM, as expected. Moreover, it is clear that ESN outperforms DNN and LSTM in MAPE by a big margin when the dynamical system that we try to model becomes more chaotic, and the data quality deteriorated.  The average times taken to train each of the models using the same data size on a GPU using a single core are also recorded. It is observed that ESN takes a much shorter time to train compared to DNN and LSTM. For example, it takes an average of 2.3 hours to train the LSTM model and an average of 6.7 minutes to train the ESN model on the same dataset. 
\begin{table}[h] 
\caption{Performance Comparison of Deep Learning Models against ESN on Data 3 \& Data 4}
\label{tab:deepvsecho}
\centering
\begin{tabular}{llcccc}
\toprule
&&\multicolumn{2}{c}{\textbf{MAPE(\%)}} &  \multicolumn{1}{c}{\textbf{Training Time}} \\
\midrule
\textbf{Model} & \textbf{Data size} &\textbf{Data 3}&\textbf{Data 4} & \textbf{(secs)} \\ 
\midrule
DNN & 100,000$\times$578 &4.19&56.54 & 1719\\  
LSTM  & 100,000$\times$578  &3.14& 52.10  & 8226 \\ 
ESN & 100,000$\times$578  &2.15& 15.96 & 407 \\ 
\bottomrule
\end{tabular}
\end{table}

\subsubsection{In-depth performance evaluation of ESN with different setups}
This set of experiments seeks to explore and investigate the effect of different ESN setups, using hyper-parameters listed in Table~\ref{tab:esnparam}, on the performance of our ESN channel models.  The experiments carried out here are itemized below.
\begin{enumerate}[label=\alph*)]
\item \textit{Using different Reservoir Initialization Methods}: As mentioned in Section~\ref{sec:reservoir}, RC is not principled enough as there is no systematic way of defining the dynamic reservoir. The reservoir is usually randomly generated at network initialization. In addition to randomly initializing the reservoir, we explored the option of using some other weight initialization approaches used in general deep learning~\cite{kumar2017weight, comparison2020,weight2021} and investigated how they affect the performance of the ESN. For the four categories of data, we ran experiments using the random, Xavier, normalized Xavier (gloriot), and HE initialization methods while keeping all other parameters constant. The model performances when each of the initialized methods were used are recorded in Table~\ref{tab:approach}. This experiment, however, was carried out without taking into consideration the echo state property.

\begin{table}[h] 
\caption{Model Performance Comparison for Different Initialization Approaches}
\label{tab:approach}
\centering
\begin{tabular}{ccccc}
\toprule
&\multicolumn{4}{c}{\textbf{MAPE(\%)}} \\
\hline
&\textbf{Random} & \textbf{Xavier} & \textbf{Norm. Xavier}& \textbf{HE} \\ 
\midrule
Data 1 & 32.54 &6.11&14.82& 9.57\\ 
Data 2  & 23.06&5.83&12.08& 10.79\\ 
Data 3 & 12.70 &2.15&6.24& 5.92\\ 
Data 4 & 46.20&15.96&27.31& 32.19\\ 
\bottomrule
\end{tabular}
\end{table}

\item \textit{Enforcing the Echo State Condition}: An investigation into the spectral radii of the matrices generated by each of the initialization approaches used above revealed, as shown in Table\ref{tab:radius}, that only the Xavier initialized reservoir matrix satisfied the echo state property at all instances. Both Gloriot and HE initialization methods have spectral radius, $\rho(\mathbf{W})$ greater than one for different reservoir sizes. In this experiment, we ensured that the echo state property is satisfied by normalizing the matrices such that the spectral radius is less than one and the results from this experiment are as recorded in Table~\ref{tab:normradius}.

\begin{table}[h]
\caption{Spectral Radius Check for different Proposed Initialization Approaches}
\label{tab:radius}
\centering
\begin{tabular}{ccccc}
\toprule
\multicolumn{4}{c}{\textbf{Spectral Radius} $\rho(\mathbf{W})$} \\
\midrule
\textbf{Reservoir Size, N} & \textbf{Xavier} & \textbf{Norm. Xavier}& \textbf{HE}  \\ 
\midrule
50 &0.63& 1.11&1.46\\
100 & 0.59& 1.03&1.41\\ 
150& 0.60& 1.05&1.49\\ 
300 & 0.61& 1.02&1.47\\ 
578 &0.60& 1.09&1.44\\ 
600& 0.58& 1.03&1.44\\ 
1200 &0.59& 1.01&1.44\\ 
2400 &0.59& 1.02&1.43\\ 
\bottomrule
\end{tabular}
\end{table}

\begin{table}[h] 
\caption{Model Performance Comparison for Different Initialization Approaches with normalized Spectral Radius [$\rho(\mathbf{W}) < 1$]}
\label{tab:normradius}
\centering
\begin{tabular}{ccccc}
\toprule
&\multicolumn{4}{c}{\textbf{MAPE(\%)}} \\
\midrule
&\textbf{Random} & \textbf{Xavier} & \textbf{Norm. Xavier}& \textbf{HE} \\ 
\midrule
Data 1 & 20.24 &6.05&6.02& 5.91\\ 
Data 2  & 14.19 &5.56&5.87& 6.41\\ 
Data 3 & 2.86 &2.16&2.17& 2.17\\ 
Data 4 & 16.2 &15.80&16.59& 16.25\\ 
\bottomrule
\end{tabular}
\end{table}

\item \textit{Using different Spectral Radii}: Here, we varied the spectral radius such that it is greater than 0 but less than 1 at an incremental step of 0.1, ensuring the echo state property is still obeyed while keeping every other parameter (the initialization method, the number of nodes in the reservoir and the activation function used) constant. The MAPE values are as recorded in Table~\ref{tab:spectral}. Figure~\ref{fig:radius} is a graphical plot of the MAPE values for different spectral radii across all the data categories.

\begin{table}[h]
\caption{Model Performance Comparison for different Spectral Radius}
\label{tab:spectral}
\centering
\begin{tabular}{ccccc}
\toprule
&\multicolumn{4}{c}{\textbf{MAPE(\%)}} \\ \midrule
$\rho(\mathbf{W})$ & \thead{Data 1} & \thead{Data 2}& \thead{Data 3} & \thead{Data 4} \\  \midrule
0.1 & 6.00& 5.98&2.15& 16.37\\ 
0.2 & 5.87& 7.49&2.15& 16.41\\ 
0.3& 6.02& 5.60&2.16& 16.35\\ 
0.4 & 6.00& 5.72&2.15& 16.64\\ 
0.5 &5.87& 7.19&2.15&15.96\\ 
0.6 & 5.75& 5.65&2.16& 16.14\\ 
0.7 & 6.04& 6.30&2.15& 16.47\\ 
0.8 & 6.06& 5.61&2.16& 16.17\\ 
0.9 & 6.02& 5.91&2.15& 16.12\\ 
\bottomrule
\end{tabular}
\end{table}

\begin{figure}[h]
	 \centering
    	 \includegraphics[width=\linewidth]{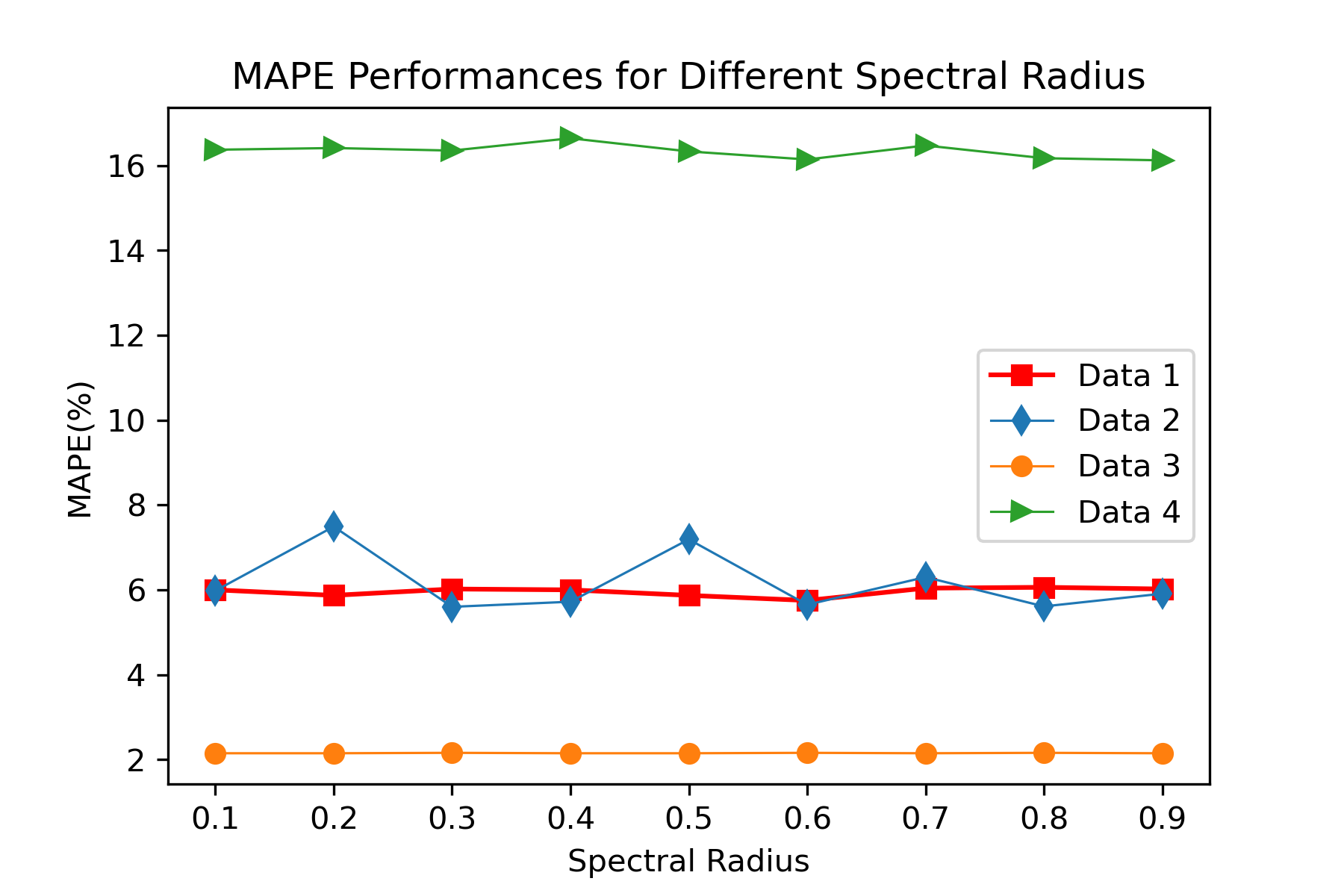}
      	\caption{MAPE Plot for Different Spectral Radius}
    \label{fig:radius}
\end{figure}

\item \textit{Using different Reservoir Sizes}: Next, we investigated the effect of varying the number of nodes in the dynamic reservoir, $\mathbf{N}$ on the performance of the model while keeping other parameters (spectral radius, initialization method, and the activation function) constant. 
Results from this experiment are as recorded in Table~\ref{tab:size}. Figure~\ref{fig:reservoirsize} shows a graphical plot of the MAPE values for the different reservoir sizes across all the data categories.

\begin{table}[h]
\caption{Model Performance Comparson for Different Sizes of the Reservoir}
\label{tab:size}
\centering
\begin{tabular}{ccccc}
\toprule
&\multicolumn{4}{c}{\textbf{MAPE(\%)}} \\
\midrule
\textbf{Reservoir Size, N} & \textbf{Data 1} & \textbf{Data 2}& \textbf{Data 3} & \textbf{Data 4} \\ 
\midrule
50 &17.02& 26.17&3.29& 16.06\\ 
100 & 10.69& 15.55&2.56& 16.10\\ 
150& 8.68& 11.91&2.32& 16.10\\ 
300 & 6.73& 7.77&2.17& 16.03\\ 
578 &5.70& 5.86&2.15& 15.96\\ 
600& 5.91& 5.97&2.21& 16.05\\ 
1200 &5.99& 6.08&2.54& 16.15\\ 
2400 &6.34& 12.13&2.87& 17.43\\ 
\bottomrule
\end{tabular}
\end{table}

\begin{figure}[h]
	 \centering
    	 \includegraphics[width=\linewidth]{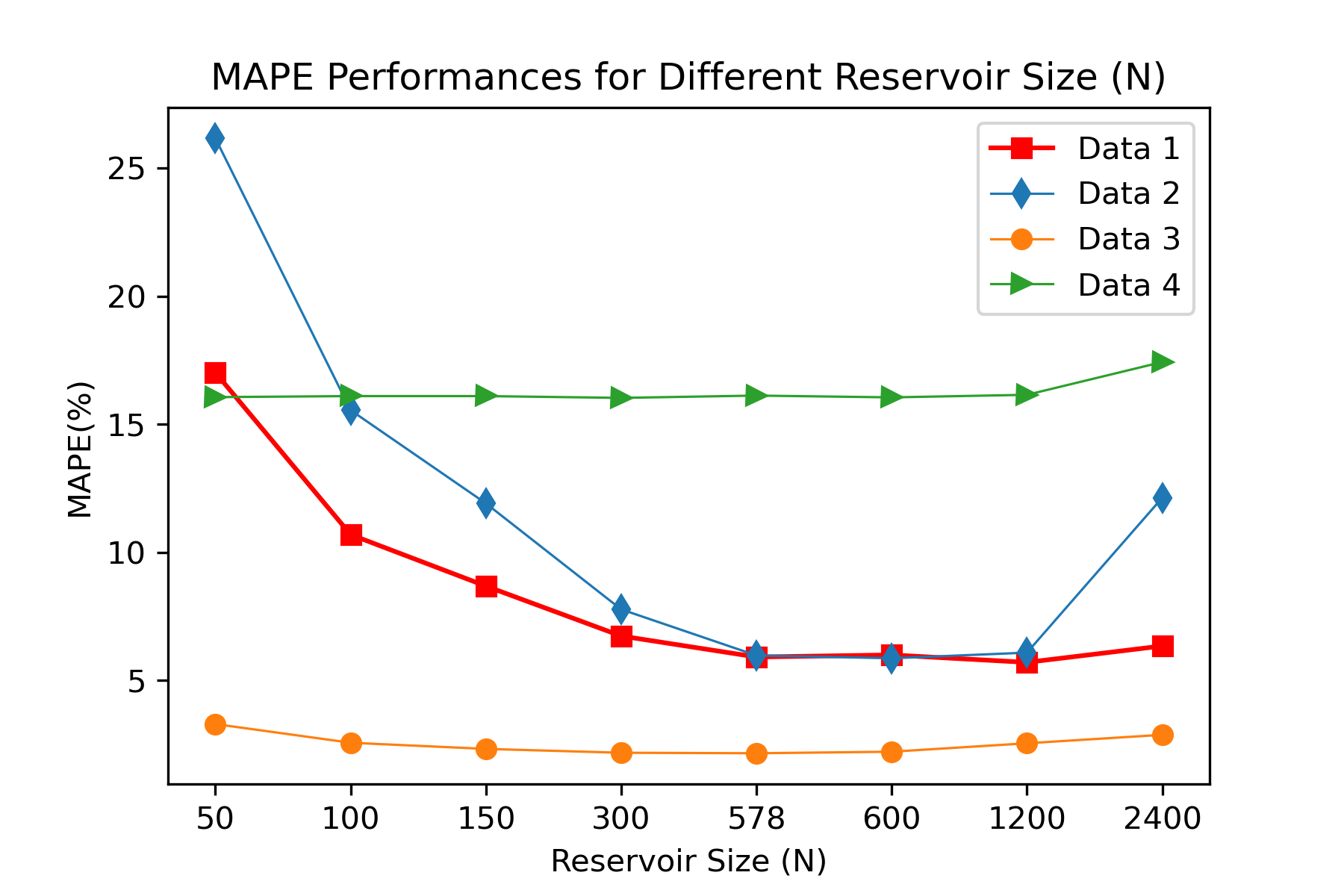}
      	\caption{MAPE Plot for Different Reservoir Sizes, N}
    \label{fig:reservoirsize}
\end{figure}

\item \textit{Using different Activation Functions}: Instead of the hyperbolic tangent ($\mathbf{tanh}$) used in the state transition equation~(\ref{eqn:four}), the effect of using some other activation functions are investigated. In addition to tanh, we also experimented with ReLU and Sigmoid activation functions. Using the Sigmoid and ReLU activation functions required normalizing the input data since the sigmoid function only gives an output between $0$ to $1$ and that of the ReLU function is between $0$ and $\infty$. Then the outputs of the activation functions are `unnormalized'. Results from these experiments are recorded in Table~\ref{tab:activation}.

\begin{table}[h]
\caption{Model Performance Comparison for Different Activation Functions}
\label{tab:activation}
\centering
\begin{tabular}{ccccc}
\toprule
&\multicolumn{4}{c}{\textbf{MAPE(\%)}} \\
\midrule
&\textbf{Data Size} &\textbf{tanh} & \textbf{ReLU} & \textbf{Sigmoid}\\ 
\midrule
Data 1 & 100,000$\times$578&6.05&6.66&7.14\\ 
Data 2 & 100,000$\times$578 & 5.56 &5.79&6.02\\ 
Data 3& 100,000$\times$578 & 2.16 &2.27&2.38\\ 
Data 4& 100,000$\times$578 & 15.96&17.21&18.11\\ 
\bottomrule
\end{tabular}
\end{table}

\item \textit{Using different Regression Models}: In minimizing error at the readout layer, the only layer where training occurs in the network, we also investigated the effect of three different regression models on the performance of the network - ridge, linear, and lasso regression, while keeping every other parameter constant. The results from these experiments are recorded in Table~\ref{tab_regr}. A sample plot of ESN predicted data is as shown in Figure~\ref{fig:predict}.
\begin{table}[ht] 
\caption{Model Performance Comparison for Different Regression Model Used at the Readout Layer}
\label{tab_regr}
\centering
\begin{tabular}{cccc}
\toprule
&\multicolumn{3}{c}{\textbf{MAPE(\%)}} \\
\midrule
&\textbf{Ridge} & \textbf{Linear} & \textbf{Lasso}\\ 
\midrule
Data 1 &5.91&6.00&5.86\\ 
Data 2  &5.89 &5.97&6.01\\ 
Data 3 & 2.15 &2.18&2.21\\ 
Data 4 & 15.96 &16.12&16.81\\ 
\bottomrule
\end{tabular}
\end{table}

\begin{figure}[ht]
	 \centering
    	 \includegraphics[width=\linewidth]{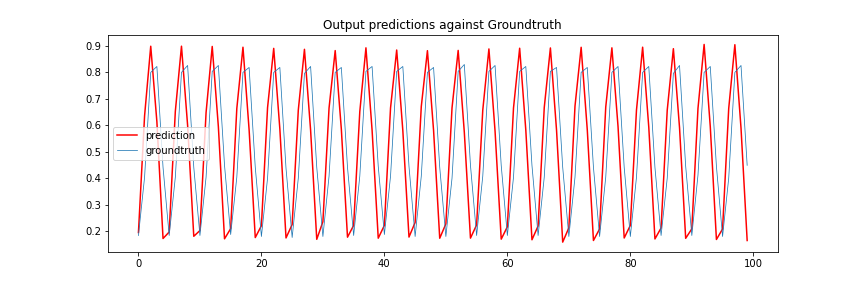}
      	\caption{Sample Plot of ESN Predicted Data}
    \label{fig:predict}
\end{figure}
\end{enumerate}

{\bf Summary of results in Section~\ref{subsec:PerformanceEvaluationESN}}: Observations from these experimental results are summarized below:
\begin{itemize}
\item ESN models perform better than the popular deep learning models such as DNN and LSTM, especially  when the data quality becomes poor as seen in Table~\ref{tab:deepvsecho} when we used the dataset with the poorest quality. It is observed that ESN takes a much shorter time to train compared to DNN and LSTM, as expected.
\item It is also observed that the echo state condition must be satisfied in order to get a good performance.  For example,  the MAPE values in Table~\ref{tab:approach} improved as seen in Table~\ref{tab:normradius} when the reservoir matrices were normalized and $\rho(W) < 1$ was enforced. 
\item RC is very robust against different spectral radii, activation functions, and regression methods used for training the output layer. For instance,  there were no significant differences in performance of the models in Tables~\ref{tab:spectral},~\ref{tab:activation} and~\ref{tab_regr} when different spectral radius, activation function and regression model respectively were used.
\item When the size of the reservoir, that is the number of nodes in the reservoir, matches the size of the input sequence into an ESN model, the model performs better.  For example, in Table~\ref{tab:size} the models gave the best performance, across all data categories, when the reservoir size was set to 578, which corresponds to the size of the sequence of the input data used in these experiments.
\end{itemize}

\subsection{Performance of the Modified ESN}
\label{subsec:PerformanceModifiedESN}
In a bid to improve the performance of the ESN, we experimented with the use of pre-trained deep learning models as a replacement for the reservoir, as highlighted in Figure~\ref{fig:esnlstm}, instead of randomized $\mathbf{W}$ using Data 3 and examine how it affects the performance of the network. Specifically, we used pre-trained DNN (PDNN) and LSTM (PLSTM), deep learning models, for these experiments, and the results are shown in Table~\ref{tab:deepecho}. 

\begin{table}[h]
\caption{Performance Comparison of Deep Learning Models, ESN and ESN with pre-trained deep models as reservoir on Data 3.}
\label{tab:deepecho}
\centering
\begin{tabular}{ccc}
\toprule
\multicolumn{3}{c}{\textbf{Data 3}}  \\
\midrule
\multicolumn{2}{c}{\textbf{Model} }& \textbf{MAPE(\%)}  \\ 
\midrule
\multicolumn{2}{l}{DNN}& 4.19\\
\multicolumn{2}{l}{LSTM }& 3.14\\
\multicolumn{2}{l}{ESN}& 2.15\\ 
\multicolumn{2}{l}{ESN + PDNN} & 3.63\\ 
\multicolumn{2}{l}{ESN + PLSTM}& 2.90\\ 
\bottomrule
\end{tabular}
\end{table}

When compared to the performances of the deep learning models,  it is observed that the ESN with pre-trained deep models as reservoir did show some improvements in performance.  There was a 13.42\% improvement in the performance of the DNN and a 7.64\% improvement in that of LSTM but did not do better than the conventional ESN model.  Thus, using pre-trained deep models as a replacement for the reservoir in the ESN might not be necessary.

\subsection{Transfer Learning Performance}
\label{subsec:TransferLearningPerformance}
Two categories of experiments were carried out to investigate if transfer learning would help improve the performance of the ESNs as stated in Section~\ref{sec:tla}. Under each category,  three  experiments were carried out. 

\subsubsection{Model trained using RF data} 
This experiment seeks to leverage trained models from the RF domain using simulated RF data and transferring them into real UWA communication domains.  RF datasets were generated using MATLAB simulation of an additive white gaussian noise (AWGN) channel.  Just as with UWA data generation,  signal bits were generated and passed through a quadrature phase-shift keying (QPSK) block before sending to a raised cosine transmit filter (RCTx) block. The output of the RCTx block was first saved and then fed to the AWGN block as the transmitted signal. The output of the AWGN channel is also saved as the received signal used for training our models.  These datasets were used in the sub-experiments itemized below.

\begin{enumerate}[label=\alph*)]
\item \textit{Training and testing models with RF data}: Deep learning models and an ESN were trained to model the AWGN channel and tested directly using solely the generated RF data. These deep models were saved and then used as the reservoir in training the readout layer of our proposed ESN architecture and also evaluated using the RF testing data. The results from these experiments are as recorded in Table~\ref{tab:rfdata}.

\begin{table}[ht]
\caption{Model Performance when models were trained and evaluated using simulated RF Data}
\label{tab:rfdata}
\centering
\begin{tabular}{llccc}
\toprule
\multicolumn{3}{c}{\textbf{RF Data}} \\
\midrule
\textbf{Model} & \textbf{Data size} & \textbf{MAPE(\%)} \\ 
\midrule
DNN & 100,000$\times$578 &6.20\\ 
LSTM & 100,000$\times$578  & 5.00\\
Echo ESN & 100,000$\times$578  & 3.50\\ 
ESN + RFP-DNN & 100,000$\times$578  & 5.34\\ 
ESN + RFP-LSTM & 100,000$\times$578  & 4.82\\ 
\bottomrule
\end{tabular}
\end{table}

\item \textit{Using RF pre-trained models and testing on Data 3}: Models from the experiment above were saved and then evaluated directly using Data 3 test dataset. The results from this experiment are as shown in Table~\ref{tab:pretrained} where RFP-DNN,  RFP-LSTM, and RFP-ESN represent RF-Pretrained DNN,  RF-Pretrained LSTM, and RF-Pretrained ESN models respectively. 
\begin{table}[ht]
\caption{Model Performance when RF-Pretrained Models from Table~\ref{tab:rfdata} were evaluated directly with Data 3.}
\label{tab:pretrained}
\centering
\begin{tabular}{llccc}
\toprule
\multicolumn{3}{c}{\textbf{Data 3}} \\
\midrule
\textbf{Model} & \textbf{Data size} & \textbf{MAPE(\%)} \\ 
\midrule
RFP-DNN & 100,000$\times$578 &17.21\\ 
RFP-LSTM & 100,000$\times$578  & 10.78\\
RFP-ESN  & 100,000$\times$578  & 68.87\\ 
ESN + RFP-DNN  & 100,000$\times$578  & 18.13\\ 
ESN + RFP-LSTM  & 100,000$\times$578  & 12.51\\ 
\bottomrule
\end{tabular}
\end{table}

\item \textit{Refining RF pre-trained models and testing on Data 3}: Next, we fine-tuned the pre-trained models from sub-experiment \textit{i} by re-training the models with Data 3 training data and re-evaluating the resultant model with Data 3 test data. Results from this experiments are as recorded in Table~\ref{tab:finetunedrf}.
\begin{table}[h] 
\caption{Model Performance when RF-Pretrained models from Table~\ref{tab:rfdata} were re-trained with Data3 and evaluated with Data 3.}
\label{tab:finetunedrf}
\centering
\begin{tabular}{llccc}
\toprule
\multicolumn{3}{c}{\textbf{Data 3}} \\
\midrule
\textbf{Model} & \textbf{Data size} & \textbf{MAPE(\%)} \\ 
\midrule
BHP-DNN & 100,000$\times$578 & 10.43\\ 
BHP-LSTM & 100,000$\times$578  & 6.32\\
BHP-ESN  & 100,000$\times$578  &  23.21\\ 
\bottomrule
\end{tabular}
\end{table}
\end{enumerate}

\subsubsection{Using Bellhop Simulated data} 
The next set of experiments seeks to import pre-trained models built from simulated datasets generated using the BELLHOP ray mathematical model~\cite{MAUACCB2017}. Although the Bellhop model is a well established mathematical model for UWA channel modeling~\cite{UACMUBRTM2017}, it failed to perform well in a real-world scenario.  When used to predict the received signals,  with real UWA data from Data 3 as the input to the model,  the average MAPE value is 57.30\%.  As with the RF data, these simulated Bellhop datasets are used in the experiments itemized below.

\begin{enumerate}[label=\alph*)]
\item \textit{Training and testing models with Bellhop data}: Deep learning models and an ESN were trained and tested using these BELLHOP generated data. The results from these experiments are as recorded in Table~\ref{tab:belldata}.
\begin{table}[h]
\caption{Model performance when models were trained and evaluated using Bellhop generated data.}
\label{tab:belldata}
\centering
\begin{tabular}{llccc}
\toprule
\multicolumn{3}{c}{\textbf{Bellhop Generated Data}} \\
\midrule
\textbf{Model} & \textbf{Data size} & \textbf{MAPE(\%)} \\ 
\midrule
DNN & 100,000$\times$578 &4.88\\ 
LSTM & 100,000$\times$578  & 3.62\\
ESN & 100,000$\times$578  & 3.14\\ 
ESN + BHP-DNN & 100,000$\times$578  & 4.02\\ 
ESN + BHP-LSTM & 100,000$\times$578  & 3.46\\ 
\bottomrule
\end{tabular}
\end{table}

\item \textit{Using Bellhop pre-trained models and testing on Data 3}: The pre-trained models, from Table~\ref{tab:belldata}, were saved and
then evaluated directly using Data 3 test data.  These deep models were saved and then used as the reservoir in training the readout layer of our proposed ESN architecture and also evaluated using the bellhop testing data. The results from these experiments are as shown in Table~\ref{tab:pretrainedbellhop} where BHP-DNN, BHP-LSTM, and BHP-ESN mean Bellhop-Pretrained DNN, Bellhop-Pretrained LSTM, and Bellhop-Pretrained ESN, respectively.
\begin{table}[h]
\caption{Model Performance when Bellhop-Pretrained models from Table~\ref{tab:belldata} are evaluated directly with Data 3. }
\label{tab:pretrainedbellhop}
\centering
\begin{tabular}{llccc}
\toprule
\multicolumn{3}{c}{\textbf{Data 3}} \\
\midrule
\textbf{Model} & \textbf{Data size} & \textbf{MAPE(\%)} \\ 
\midrule
BHP-DNN & 100,000$\times$578 & 6.26\\ 
BHP-LSTM & 100,000$\times$578  & 4.92\\
BHP-ESN  & 100,000$\times$578  & 3.98\\ 
ESN + BHP-DNN  & 100,000$\times$578  & 5.81\\ 
ESN + BHP-LSTM  & 100,000$\times$578  & 4.10\\ 
\bottomrule
\end{tabular}
\end{table}

\item \textit{Refining Bellhop pre-trained models and testing on Data 3}: Pre-trained models  from Table~\ref{tab:belldata} were fine-tuned by re-training the models with Data 3 training data and re-evaluating the resultant model with Data 3. Results from these experiments are as recorded in Table~\ref{tab:finetunedbellhop}.
\begin{table}[h]
\caption{Model Performance when Bellhop-Pretrained models from Table~\ref{tab:belldata} were re-trained with Data 3 and evaluated with Data 3 testing data.}
\label{tab:finetunedbellhop}
\centering
\begin{tabular}{llccc}
\toprule
\multicolumn{3}{c}{\textbf{Data 3}} \\
\midrule
\textbf{Model} & \textbf{Data size} & \textbf{MAPE(\%)} \\ 
\midrule
BHP-DNN & 100,000$\times$578 & 4.91\\ 
BHP-LSTM & 100,000$\times$578  & 3.23\\
BHP-ESN  & 100,000$\times$578  & 2.51\\ 
\bottomrule
\end{tabular}
\end{table}
\end{enumerate}

{\bf Summary of Results of Section~\ref{subsec:TransferLearningPerformance}}: 
\begin{itemize}
\item Transfer learning does not perform well when transferring models trained with very different data or domains.  For example,  using models built with simulated RF data and transferring directly to underwater data did not give a good result as observed in Table~\ref{tab:pretrained}. 
\item On the other hand,  transfer learning tends to perform reasonably well when the base domain is related to the target domain.  For instance,  when models trained with simulated Bellhop data were transferred and evaluated directly on underwater data,  there was a notable improvement in performance as seen in Table~\ref{tab:pretrainedbellhop}.
\item Transfer learning can be further improved by refining or fine-tuning the pre-trained base model with the target datasets.  When both RF pre-trained and Bellhop pre-trained models were retrained using the UWA datasets, improvements in the performance of the models are observed in Tables~\ref{tab:finetunedrf} and~\ref{tab:finetunedbellhop}.
\end{itemize}

\section{Conclusion}
\label{sec:concl}
Two key objectives were pursued in this work. The first objective is to mitigate the approximations and unrealistic assumptions made by the mathematical models for UWA channel models by providing a data-driven approach to UWA channel modeling.  To achieve this first goal, data generation and collection experiments were carried out in a water tank and in Lake Tahoe under different levels of disturbances. Then the obtained datasets were used to train deep learning models (DNN and LSTM) as well as the ESN to obtain models with high fidelity in a real-world scenario. It is observed that ESN performs better than DNN and LSTM in terms of prediction accuracy and has much less computational cost in terms of training time. The performance gap becomes larger when the data quality gets worse (more chaotic data), which is in agreement with previous studies in the literature that RC is very effective in modeling chaotic dynamical systems. 
 The second objective is to examine how RC performs under various setups including different initialization methods of the reservoir, various spectral radii,  the reservoir size,  the activation function used, and the regression method used for training the output layer. It is shown that ESN is quite robust to these different settings as long as the spectral radius satisfy the echo state property.  It is also observed that when the size of the reservoir matches the size of the input sequence, the ESN model performs better. Furthermore, the performance of transfer learning in ESN is also evaluated. Although Bellhop mathematical model has a very poor performance by itself, when ESN models trained with simulated Bellhop data were transferred and evaluated on real-world underwater data,  there was a noticeable improvement in ESN performance.
 Lastly, a modified ESN implementation was examined where the reservoir of the ESN is replaced with a pre-trained deep learning model. Though this approach gives a slightly better performance when compared to pre-trained deep learning models, it is not as efficient as the conventional ESN. 

In this study, the trained ESN model,  using real-world UWA communications data,  is a data-driven black box model that performs sequence-to-sequence or point-to-point prediction very well with time-series I/Q data as the input to the model.  It could be used to carry out large scale simulations of UWA communications or obtain a large amount of high fidelity data when it is difficult to obtain that kind of data from physical underwater data collections. While it provides a more realistic model when compared to mathematical models, it may also be over specified due to the peculiarities of the measurement environment. It could also be generalized to model a network setting with multiple transceivers or could be refined to model the ocean environment or any other environment with time-series data.  For instance, we can generalize the developed channel model to a network setting with multiple transmitting-receiving (Tx/Rx) pairs, as long as the multiple Tx/Rx pairs are orthogonal. If we consider multiple simultaneous transmissions in non-orthogonal UWA channels, new channel model must be trained based on new datasets that capture the mutual interference. The model can also be trained in the reverse direction to recover the transmitted signals based on the received signals, which may be an alternative implementation for the  software-based receiver. This is one of our future efforts.

\section*{Acknowledgment}
\label{sec:acknowledgement}
This research work is supported in part by the U.S. Army Research Office (ARO) under agreement number W911NF-20-2-0266 and the U.S. Office of the Under Secretary of Defense for Research and Engineering (OUSD(R\&E)) under agreement number FA8750-15-2-0119. The U.S. Government is authorized to reproduce and distribute reprints for governmental purposes notwithstanding any copyright notation thereon. The views and conclusions contained herein are those of the authors and should not be interpreted as necessarily representing the official policies or endorsements, either expressed or implied, of the U.S. Army Research Office (ARO) or the Office of the Under Secretary of Defense for Research and Engineering (OUSD(R\&E)) or the U.S. Government.

\bibliographystyle{IEEEtran}
\bibliography{references}

\pagebreak

\begin{IEEEbiography}[{\includegraphics[width=1in,height=1.25in,clip,keepaspectratio]{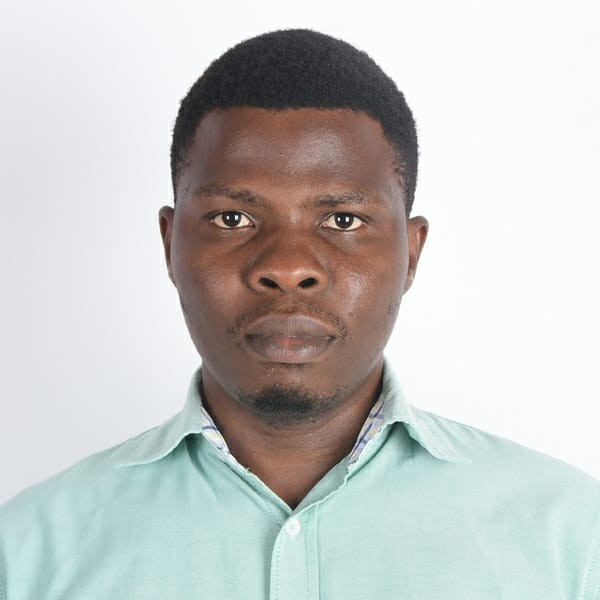}}]{Oluwaseyi Onasami} was born in Nigeria, West Africa. He received his Bachelor's degree in Electronic and Electrical Engineering at the Obafemi Awolowo University,  Nigeria in 2010 and is currently pursuing his masters degree in Electrical and Computer Engineering at the U.S. DOD Center of Excellence in Research and Education for Big Military Data Intelligence (CREDIT Center),  Department of Electrical and Computer Engineering, Prairie View A\&M University, Texas, USA. His research interests are in the area of big data processing, deep learning and applied artificial intelligence.
\end{IEEEbiography}

\begin{IEEEbiography}[{\includegraphics[width=1in,height=1.25in,clip,keepaspectratio]{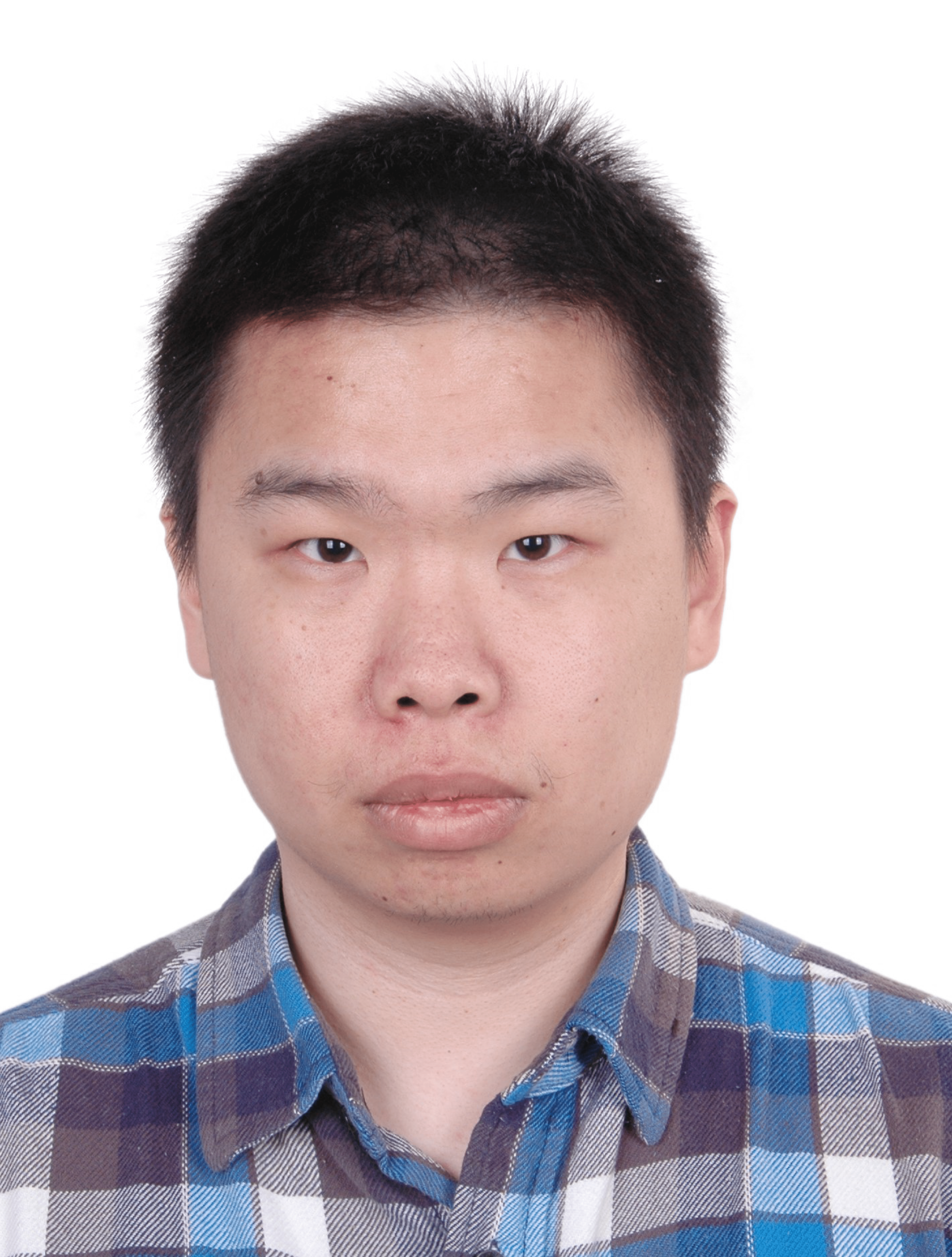}}]{Ming Feng} received his Bachelor’s degree in communication engineering from Taiyuan University of Technology in 2011 and received M.S. degree in Electrical Engineering from Stevens Institute of Technology in 2016. He received his Ph.D. degree in Electrical and Biomedical Engineering at the University of Nevada, Reno in 2020. Currently, he worked as a research engineer in Hughes Network Systems, LLC. His research interests are in optimal resource allocation with application complex wireless networks, UAV-assisted next generation networks, etc.
\end{IEEEbiography}

\begin{IEEEbiography}[{\includegraphics[width=1in,height=1.25in,clip,keepaspectratio]{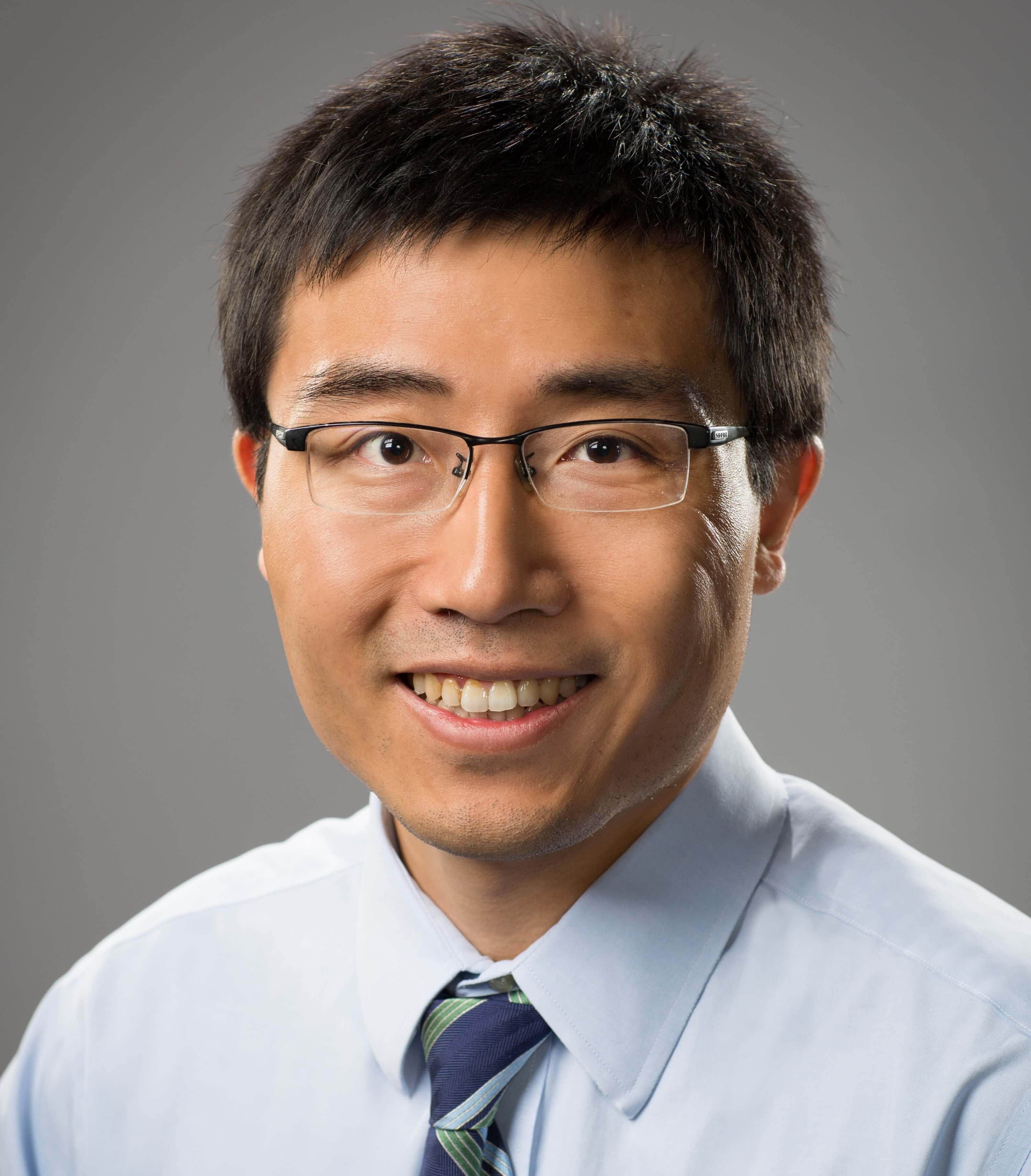}}]{Hao Xu } received his Master’s degree in Electrical Engineering from Southeast University in 2009, and his Ph.D. degree from the Missouri University of Science and Technology (formerly, the University of Missouri-Rolla), Rolla in 2012. He currently holds an assistant professor position with the Department of Electrical and Biomedical Engineering at the University of Nevada, Reno. His research interests are in AI-based optimization for the wireless network, underwater network, and reliable communication network development.
\end{IEEEbiography}

\begin{IEEEbiography}[{\includegraphics[width=1in,height=1.25in,clip,keepaspectratio]{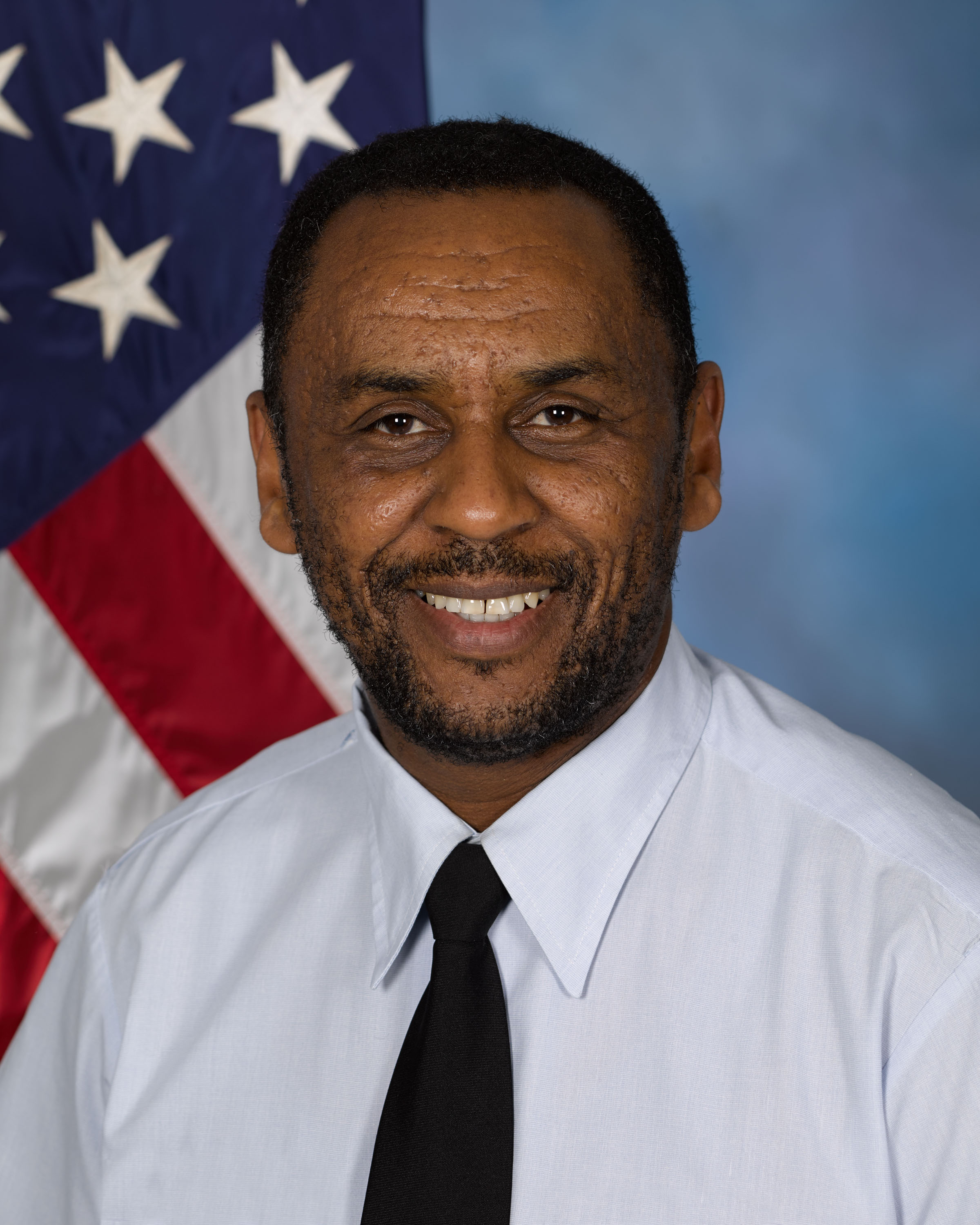}}]{Mulugeta Haile} is a senior aerospace research engineer and a research team leader at the U.S. Army Research Laboratory (ARL), Aberdeen Proving Ground, Maryland. He received his graduate degrees in Electrical and Aerospace Engineering from the University of Florida, Gainesville.  Dr. Haile has led several research projects in aerospace mechanics and artificial intelligence (AI). He is the founder of the intelligent mechanics (iMechanics) lab at ARL. He has authored several research papers in top journals, conference proceedings, and book chapters and has over 23 years of research and development experience. His current research interests are in embodied intelligence, scientific machine learning, and AI-assisted creativity.
\end{IEEEbiography}

\begin{IEEEbiography}[{\includegraphics[width=1in,height=1.25in,clip,keepaspectratio]{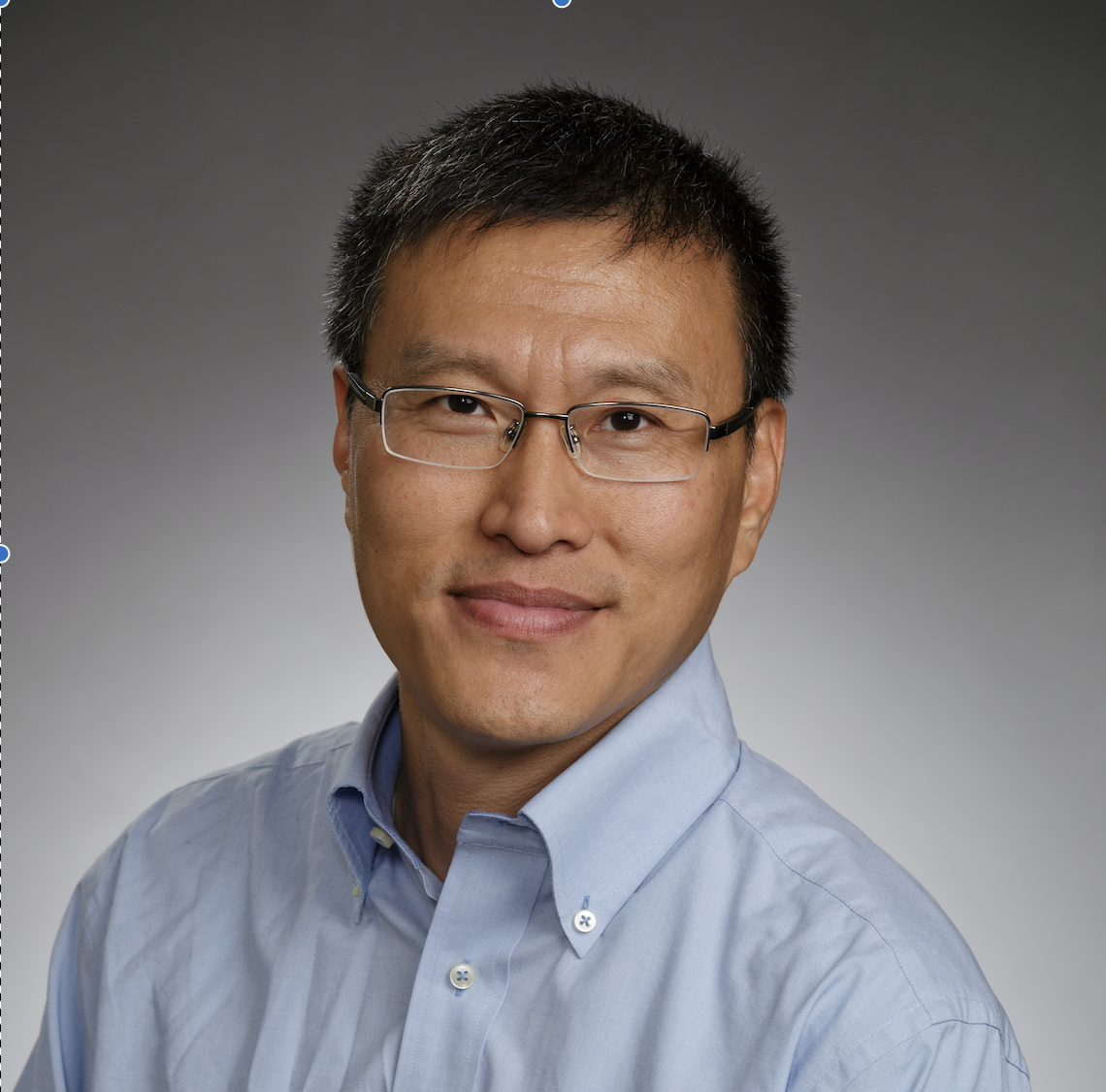}}]{Lijun Qian} (SM'08) is Regents Professor and holds the AT\&T Endowment in the Department of Electrical and Computer Engineering at Prairie View A\&M University (PVAMU), a member of the Texas A\&M University System, Prairie View, Texas, USA. He is also the Director of the Center of Excellence in Research and Education for Big Military Data Intelligence (CREDIT Center). He received BS from Tsinghua University, MS from Technion-Israel Institute of Technology, and PhD from Rutgers University. Before joining PVAMU, he was a member of technical staff of Bell-Labs Research at Murray Hill, New Jersey. He was a visiting professor of Aalto University, Finland. His research interests are in the area of big data processing, artificial intelligence, wireless communications and mobile networks, network security and intrusion detection, and computational and systems biology.
\end{IEEEbiography}

\EOD

\end{document}